\documentclass[longauth,bibyear]{aa}

\usepackage{graphicx}
\usepackage{txfonts}
\usepackage{lipsum}
\usepackage{subcaption}         
\usepackage{lscape}            
\usepackage{placeins}

\usepackage{natbib}
\bibpunct{(}{)}{;}{a}{}{,} 
\usepackage[colorlinks=true]{hyperref}
\hypersetup{colorlinks=true,allcolors=blue,citecolor=blue}
\usepackage{xcolor}
\usepackage{amssymb}
\usepackage{amsmath}

\usepackage{gensymb}

\RequirePackage[normalem]{ulem} 
\RequirePackage{xcolor}\definecolor{brickred}{rgb}{0.80,0.08,0.12}\definecolor{wrongultramarine}{rgb}{0.07,0.04,0.56}

\defcitealias{agolRefiningTransittimingPhotometric2021}{A21}
\defcitealias{jahandarChemicalFingerprintsDwarfs2025}{Jahandar el al. 2025}

\begin{document}

   \title{Radial velocity detection of the \hbox{TRAPPIST-1} planetary system with SPIRou and NIRPS}

\author{
Alexandrine~L'Heureux\inst{1}\corrauth{alexandrine.lheureux@umontreal.ca},
Ren\'e~Doyon\inst{1,2}\email{rene.doyon@umontreal.ca},
Charles~Cadieux\inst{1,3}\email{charles.cadieux@unige.ch},
Pascal~Petit\inst{4}\email{ppetit@irap.omp.eu},
Olivia~Lim\inst{1}\email{olivia.lim@umontreal.ca},
\'Etienne~Artigau\inst{1,2}\email{etienne.artigau@umontreal.ca},
Neil~J.~Cook\inst{1}\email{neil.cook@umontreal.ca},
Eric~Agol\inst{5}\email{agol@uw.edu},
Xavier~Bonfils\inst{6}\email{xavier.bonfils@cnrs.fr},
Julien~Morin\inst{7}\email{julien.morin@umontpellier.fr},
Stefano~Bellotti\inst{8}\email{bellotti@strw.leidenuniv.nl},
Claire~Moutou\inst{4}\email{claire.moutou@irap.omp.eu},
Leslie~Moranta\inst{1,9}\email{leslie.moranta@umontreal.ca},
Jean-Fran\c{c}ois~Donati\inst{4}\email{jean-francois.donati@irap.omp.eu},
Luc~Arnold\inst{10}\email{larnold@cfht.hawaii.edu},
Xavier~Delfosse\inst{6}\email{Xavier.Delfosse@univ-grenoble-alpes.fr},
Guillaume~H\'ebrard\inst{11}\email{hebrard@iap.fr},
Romain~Allart\inst{1}\email{romain.allart@umontreal.ca},
Fran\c{c}ois~Bouchy\inst{3}\email{Francois.Bouchy@unige.ch},
Lucile~Mignon\inst{3,6}\email{lucile.mignon@univ-grenoble-alpes.fr},
Izan~de~Castro~Le\~ao\inst{12}\email{izan@fisica.ufrn.br},
Dany~Mounzer\inst{3}\email{Dany.Mounzer@unige.ch},
Khaled~Al~Moulla\inst{13,3}\email{khaled.almoulla@astro.up.pt},
Fr\'ed\'erique~Baron\inst{1,2}\email{frederique.baron@umontreal.ca},
Bj\"orn~Benneke\inst{14,1}\email{bjorn.benneke@umontreal.ca},
Mathis~Bouffard\inst{1}\email{mathis.bouffard@umontreal.ca},
Casey~Brinkman\inst{15}\email{casey.brinkman@mcgill.ca},
Bruno~L.~Canto~Martins\inst{12}\email{brunocanto@fisica.ufrn.br},
Andr\'es~Carmona\inst{6}\email{andres.carmona.astronomy@gmail.com},
Ryan~Cloutier\inst{16}\email{ryan.cloutier@mcmaster.ca},
Marion~Cointepas\inst{3,6}\email{marion.cointepas@unige.ch},
Nicolas~B.~Cowan\inst{15,17}\email{nicolas.cowan@mcgill.ca},
Eduardo~Cristo\inst{13}\email{Eduardo.cristo@astro.up.pt},
Roseane~de~Lima~Gomes\inst{1,12}\email{roseanelima@fisica.ufrn.br},
Jos\'e~Renan~De~Medeiros\inst{12}\email{renan@fisica.ufrn.br},
Xavier~Dumusque\inst{3}\email{xavier.dumusque@unige.ch},
Dasaev~O.~Fontinele\inst{12}\email{dasaev.fontinele.604@ufrn.edu.br},
Thierry~Forveille\inst{6}\email{Thierry.Forveille@univ-grenoble-alpes.fr},
Yolanda~G.~C.~Frensch\inst{13}\email{yolanda.frensch@astro.up.pt},
Jonathan~Gagn\'e\inst{9,1}\email{jonathan.gagne.1@gmail.com},
Jonay~I.~Gonz\'alez~Hern\'andez\inst{18,19}\email{jonay.gonzalez@iac.es},
Nicole~Gromek\inst{16}\email{gromekn@mcmaster.ca},
Melissa~J.~Hobson\inst{3}\email{melissa.hobson@unige.ch},
Vigneshwaran~Krishnamurthy\inst{15}\email{vigneshwaran.krishnamurthy@mcgill.ca},
Pierrot~Lamontagne\inst{20,1,}\email{lamontagne@iaa.es},
Lison~Malo\inst{1,2}\email{lison.malo@umontreal.ca},
Eder~Martioli\inst{12,21}\email{emartioli@lna.br},
Yuri~S.~Messias\inst{1,12}\email{yuri.messias1@gmail.com},
Louise~D.~Nielsen\inst{3,22,23}\email{Louise.Nielsen@lmu.de},
Ares~Osborn\inst{6,16,24}\email{dr.ares.osborn@gmail.com},
L\'ena~Parc\inst{3}\email{lena.parc@unige.ch},
Caroline~Piaulet-Ghorayeb\inst{25,1}\email{carolinepiaulet@uchicago.edu},
Nuno~C.~Santos\inst{13,26}\email{nuno.santos@astro.up.pt},
Bennett~Neil~Skinner\inst{16,27,}\email{skinnb1@mcmaster.ca},
Avidaan~Srivastava\inst{1}\email{avidaan.srivastava@umontreal.ca},
Atanas~K.~Stefanov\inst{18,19}\email{atanas.stefanov@iac.es},
Alejandro~Su\'arez~Mascare\~no\inst{18,19}\email{asm@iac.es},
Gregg~Wade\inst{28,29}\email{gregg.wade@rmc.ca},
Joost~P.~Wardenier\inst{30,1}\email{joost.wardenier@unibe.ch},
Drew~Weisserman\inst{16}\email{weisserd@mcmaster.ca}
}

\institute{
\inst{1}Institut Trottier de recherche sur les exoplan\`etes, D\'epartement de Physique, Universit\'e de Montr\'eal, 1375 Ave Th\'er\`ese-Lavoie-Roux, Montr\'eal, QC, H2V 0B3, Canada\\
\inst{2}Observatoire du Mont-M\'egantic, Qu\'ebec, Canada\\
\inst{3}Observatoire de Gen\`eve, D\'epartement d’Astronomie, Universit\'e de Gen\`eve, Chemin Pegasi 51, 1290 Versoix, Switzerland\\
\inst{4}CNRS, OMP, Universit\'e de Toulouse, 14 Avenue Belin, 31400 Toulouse, France\\
\inst{5}Department of Astronomy, University of Washington, Seattle, WA 98195, USA\\
\inst{6}Universit\'e Grenoble Alpes, CNRS, IPAG, 38000 Grenoble, France\\
\inst{7}LUPM, Universit\'e de Montpellier, CNRS, 34095 Montpellier, France\\
\inst{8}Sterrewacht Leiden, Universiteit Leiden, PO Box 9513, 2300 RA Leiden, Netherlands\\
\inst{9}Plan\'etarium de Montr\'eal, Espace pour la Vie, 4801 Ave Pierre-de Coubertin, Montr\'eal, QC, H1V 3V4, Canada\\
\inst{10}Canada-France-Hawai\okina i Telescope, CNRS, Kamuela, HI 96743, USA\\
\inst{11}Institut d’astrophysique de Paris, UMR7095 CNRS, Sorbonne Universit\'e, 98bis boulevard Arago, 75014 Paris, France\\
\inst{12}Departamento de F\'isica Te\'orica e Experimental, Universidade Federal do Rio Grande do Norte, Campus Universit\'ario, Natal, RN, 59072-970, Brazil\\
\inst{13}Instituto de Astrof\'isica e Ci\^encias do Espa\c{c}o, Universidade do Porto, CAUP, Rua das Estrelas, 4150-762 Porto, Portugal\\
\inst{14}Department of Earth, Planetary, and Space Sciences, University of California, Los Angeles, CA 90095, USA\\
\inst{15}Department of Physics, McGill University, 3600 rue University, Montr\'eal, QC, H3A 2T8, Canada\\
\inst{16}Department of Physics \& Astronomy, McMaster University, 1280 Main St W, Hamilton, ON, L8S 4L8, Canada\\
\inst{17}Department of Earth \& Planetary Sciences, McGill University, 3450 rue University, Montr\'eal, QC, H3A 0E8, Canada\\
\inst{18}Instituto de Astrof\'isica de Canarias (IAC), Calle V\'ia L\'actea s/n, 38205 La Laguna, Tenerife, Spain\\
\inst{19}Departamento de Astrof\'isica, Universidad de La Laguna (ULL), 38206 La Laguna, Tenerife, Spain\\
\inst{20}Instituto de Astrof\'isica de Andaluc\'ia (IAA-CSIC), Glorieta de la Astronom\'ia s/n, 18008 Granada, Spain\\
\inst{21}Laborat\'orio Nacional de Astrof\'isica, Rua Estados Unidos 154, 37504-364, Itajub\'a - MG, Brazil\\
\inst{22}European Southern Observatory (ESO), Karl-Schwarzschild-Stra{\ss}e 2, 85748 Garching bei M\"unchen, Germany\\
\inst{23}Universit\"atssternwarte, Fakult\"at f\"ur Physik, Ludwig-Maximilians-Universit\"at M\"unchen, Scheinerstra{\ss}e 1, 81679 M\"unchen, Germany\\
\inst{24}Department of Physics, The University of Warwick, Gibbet Hill Road, Coventry, CV4 7AL, United Kingdom\\
\inst{25}Department of Astronomy \& Astrophysics, University of Chicago, 5640 South Ellis Avenue, Chicago, IL 60637, USA\\
\inst{26}Departamento de F\'isica e Astronomia, Faculdade de Ci\^encias, Universidade do Porto, Rua do Campo Alegre, 4169-007 Porto, Portugal\\
\inst{27}Origins Institute, McMaster University, 1280 Main St W, Hamilton, ON, L8S 4L8, Canada\\
\inst{28}Department of Physics, Engineering Physics, and Astronomy, Queen’s University, 99 University Avenue, Kingston, ON K7L 3N6, Canada\\
\inst{29}Department of Physics and Space Science, Royal Military College of Canada, 13 General Crerar Cres., Kingston, ON K7P 2M3, Canada\\
\inst{30}Space Research and Planetary Sciences, Physikalisches Institut, Universit\"at Bern, Gesellschaftsstrasse 6, 3012 Bern, Switzerland\\
}

   \date{Received 17 March 2026 / Accepted 17 August 2026}

  \abstract 
   {The \hbox{TRAPPIST-1} system is well-known for its seven transiting Earth-sized exoplanets. It has been extensively studied and characterized, notably with transit timing variations (TTVs) to precisely measure the mass of the planets. Using near-infrared spectroscopic observations obtained as part of the SPIRou Legacy Survey and the NIRPS Guaranteed Time Observation programs, we aimed to verify those values through radial velocity (RV) measurements of the system. Our RV analysis reveals that the current data do not have the precision required to individually detect the \hbox{TRAPPIST-1} planets. However, we confidently detect ($\Delta\ln\mathcal{Z}=7.53$, 1860:1 odds) the combined RV signature of the planets by informing their relative masses on the TTV analysis, with TRAPPIST-1\,b as a proxy of the whole system. For the first time, the RV signal of the \hbox{TRAPPIST-1} system is recovered: we find a RV semi-amplitude of $K_{\mathrm{b}}=3.65^{+0.78}_{-0.83}$\,m\,s$^{-1}$ corresponding to a planetary mass of $M_{p,\,\mathrm{b}}=1.31\pm0.29$\,M$_\oplus$, demonstrating that the RV measurements are consistent with the TTV model ($M_{p,\,\mathrm{b};\,\text{TTV}}=1.374\pm0.069$\,M$_\oplus$). Additionally, the NIRPS RVs constrain the presence of giant planets beyond the snow line, excluding Saturn-mass planets out to 2.7-yr orbits and Neptune-mass objects out to 20~d. Through RV, we determined the stellar activity period to be of $3.22^{+0.22}_{-0.20}$~d. Its agreement with photometric measurements (\textit{K2} and \textit{TESS}) confirms stellar rotation as the origin of the $\sim3.3$-d periodicity observed for TRAPPIST-1. We further investigated stellar activity with SPIRou polarimetric measurements, placing an upper limit on the longitudinal field ($|B_l|<40$\,G, $3\sigma$). This limit is compatible with a weak multipolar large-scale magnetic geometry, as observed in some of the later-type rapidly rotating M dwarfs.}


   \authorrunning{A. L'Heureux et al.}
   \titlerunning{RV Detection of the \hbox{TRAPPIST-1} Planetary System with SPIRou and NIRPS}

   \maketitle
   \nolinenumbers

\section{Introduction} \label{sec:introduction}

The \hbox{TRAPPIST-1} system has been under close scrutiny since its initial discovery by \citet{gillonTemperateEarthsizedPlanets2016}. The system features seven transiting Earth-sized exoplanets \citep{gillonSevenTemperateTerrestrial2017}, on short orbits \citep[$P<19$~d;][]{lugerSevenplanetResonantChain2017} around a nearby  \citep[$d\approx12.5$\,pc;][]{gaiacollaborationGaiaDataRelease2023} M8$\pm0.5$V  star also known as 2MUCD~12171 \citep{gizisNewNeighbors2MASS2000, liebertRIPhotometry2MASSselected2006, schmidtActivityKinematicsUltracool2007}. These planets, of which three are in the conservative habitable zone, have large radius and mass ratios in relation to their small host star \citep[$R_\star\sim0.12\,\text{R}_\odot$ and $M_\star\sim0.09\,\text{M}_\odot$;][]{ducrotTRAPPIST1GlobalResults2020}, allowing for a detailed analysis of the system. These favorable conditions make \hbox{TRAPPIST-1} an ideal target for the characterization of terrestrial exoplanets.

Notably, the planetary masses were measured to remarkable precisions of 3--5\% by \citet{agolRefiningTransittimingPhotometric2021} (hereafter \citetalias{agolRefiningTransittimingPhotometric2021}) using transit-timing variations (TTVs), a method based on measuring the variation of the mid-transit time due to the mutual gravitational perturbations of the planets \citep{agolDetectingTerrestrialPlanets2005,holmanUseTransitTiming2005}. While the mass of a planet can be inferred from TTV measurements, it is degenerate with its orbital eccentricity \citep{lithwickExtractingPlanetMass2012, dawsonTOI216bTOI216Two2019}. To reduce this effect, \citetalias{agolRefiningTransittimingPhotometric2021} used the ``chopping'' variations \citep[short timescale TTVs;][]{deckMeasurementPlanetMasses2015} detected for each \hbox{TRAPPIST-1} planet with the exception of d, and transit data covering more than twice the resonant timescale of the TTVs \citep[$P_{\text{TTV}}$;][]{lithwickExtractingPlanetMass2012}. Moreover, all the \hbox{TRAPPIST-1} planets are expected to circularize ($e<0.01$) within a few million years, even assuming a dissipation factor as low as 10\% that of the Earth \citep{lugerSevenplanetResonantChain2017}. With estimates placing the system's age at more than 1\,Gyr \citep{burgasserAgeTRAPPIST1System2017}, low-eccentricity values can confidently be assumed for the TTV model of TRAPPIST-1. 

Multiple studies have shown a tendency towards lower densities for planets characterized by TTVs rather than radial velocity (RV) measurements \citep[e.g.,][]{wuDensityEccentricityKepler2013, weissMASSRADIUSRELATION2014, millsPlanetaryMassradiusRelation2017, adibekyanDensityDiscrepancyTransittiming2024}. This trend can be partially explained by the detection bias associated with each method: RV is sensitive to more massive planets and transit (thus TTVs), to larger ones. \citet{leleuRemovingBiasesDensity2023} showed that a portion of this discrepancy can be resolved at the TTV extraction level, replacing the classical approach of fitting the individual mid-transit times by a photo-dynamical fit of the light curve. This method yields TTVs of higher amplitude, meaning planets of higher mass. To obtain the planetary masses of \hbox{TRAPPIST-1}, \citetalias{agolRefiningTransittimingPhotometric2021} used the classical approach, only including a photo-dynamical fit to improve the planetary radius ratios and the stellar density. This was a practical choice, informed by the large number of free parameters required by a fully photo-dynamical model of the \hbox{TRAPPIST-1} system. In such a complex model, sampling the parameter space is a challenge and would require significant computational resources. Therefore, the masses of the \hbox{TRAPPIST-1} planets might be slightly affected by the TTV extraction method.

RV measurements are complementary to any TTV analysis. While being sensitive to non-transiting planets, the additional constraints also help break the mass-eccentricity degeneracy \citep[e.g.,][]{trifonovPairWarmGiant2021, trifonovTOI2525PairMassive2023, dawsonPreciseTransitRadialvelocity2021} and derive absolute masses for the objects in the system \citep{agolDetectingTerrestrialPlanets2005, almenaraAbsoluteMassesRadii2015}. In RV, we can probe longer and non-resonant orbits, constraining the presence of planets TTVs might be blind to. By combining RV measurements and TTVs, we get a more detailed picture of the system. A TTV+RV analysis has yet to be done for \hbox{TRAPPIST-1} and would shed light on the reliability of its TTV masses while also probing for the presence of additional non-transiting planets. 

An RV detection is even more critical in the context of the study of \citet{millsPlanetaryMassradiusRelation2017}, who compiled all planets with a non-zero mass measurement at the $2\sigma$ level published for both the RV and TTV methods independently. Out of their sample of nine planets, they found an overall good agreement (1--2$\sigma$ level) with the exception of Kepler--89\,d, which differs at the $4\sigma$ level \citep{weissMassKOI94dRelation2013, masudaCharacterizationKOI94System2013, jontof-hutterTESSObservationsKepler2022}. This discrepancy demonstrates the importance of using different datasets and methods as validation. Indeed, the disagreement is beyond the mass-eccentricity degeneracy: the TTV solutions of the Kepler--89 system favor small eccentricities \citep[$e<0.1$;][]{masudaCharacterizationKOI94System2013, jontof-hutterTESSObservationsKepler2022}, but are inconsistent with the RV solution whether it is circular or eccentric \citep{weissMassKOI94dRelation2013}.

Characterizing the masses of the \hbox{TRAPPIST-1} planets with RV requires $\sim$1\,m\,s$^{-1}$ precision. Reaching this precision represents a serious challenge for current facilities as late-type M dwarfs such as \hbox{TRAPPIST-1} are generally too faint ($V>18$\,mag) for state-of-the-art optical spectrographs such as HARPS \citep{mayorSettingNewStandards2003}, ESPRESSO \citep{pepeESPRESSOVLTOnsky2021}, EXPRES \citep{jurgensonEXPRESNextGeneration2016}, and MAROON-X \citep{seifahrtMAROONXRadialVelocity2018}. Moreover, M dwarfs are often magnetically active
\citep{noyesRotationConvectionMagnetic1984, donatiLargescaleMagneticTopologies2008, morinLargescaleMagneticTopologies2010, newtonHaEMISSIONNEARBY2017, reinersMagnetismRotationNonthermal2022}. Their stellar surface is heterogeneous, made up of spots and faculae that move into our line of sight due to stellar rotation, generating signals in the RV time series that could result in the detection of a fake planet and hinder our ability to detect one that is real \citep[e.g.,][]{quelozNoPlanetHD2001, bonfilsHARPSSearchSouthern2007, desortSearchExoplanetsRadialvelocity2007, borgnietUsingSunEstimate2015, meunierUsingSunEstimate2015, suarezmascarenoRotationPeriodsLatetype2015, carmonaNearIROpticalRadial2023}. Stellar variability is expected to be particularly challenging for \hbox{TRAPPIST-1} RV measurements, with periodicity and amplitude comparable to that of the planets \citep[$\pm5$\,m\,s$^{-1}$;][]{kleinSimulatingRadialVelocity2019}. Near-infrared (NIR) precision spectrographs such as SPIRou \citep[SpectroPolarimètre InfraRouge;][]{donatiSPIRouNIRVelocimetry2020}, NIRPS \citep[Near-InfraRed Planet Searcher;][]{bouchyNIRPSJoiningHARPS2025}, CARMENES
\citep{quirrenbachCARMENESHighresolutionSpectra2018}, and HPF \citep[Habitable-zone Planet Finder;][]{mahadevanHabitablezonePlanetFinder2012, mahadevanHabitablezonePlanetFinder2014}, were specially designed to explore the population of planetary systems around low-mass dwarf stars. At longer wavelengths, they appear brighter and, for most stars, the effects of stellar activity can be mitigated to a greater extent \citep[e.g.,][]{carmonaNearIROpticalRadial2023} due to (1) spots and faculae having a lower brightness contrast with the photosphere \citep[e.g.,][]{huelamoTWHydraeEvidence2008, mahmudStarspotinducedOpticalInfrared2011, baileyPRECISEINFRAREDRADIAL2012}, and (2) a variation in the contribution to the depth of absorption lines between the spots and the photosphere \citep{larueChromaticityStellarActivity2025}. This combination of effects can lead to a significant reduction of the RV signature of stellar heterogeneities in the NIR. Still, the effect is not always straightforward, with  magnetic stars showcasing significantly more Zeeman splitting at longer wavelengths, affecting RV measurements \citep{reinersRadialVelocitySignatures2013}.

This paper presents a comprehensive analysis of RV measurements obtained with the NIR spectrographs SPIRou and NIRPS, along with a reanalysis of the \textit{K2} and \textit{TESS} photometry to re-derive the rotation period of TRAPPIST-1. We show that the RV data are sensitive to both the stellar activity, constraining the rotation period of the star, and the planetary signature of the \hbox{TRAPPIST-1} system, yielding masses consistent with TTV measurements. The paper is structured as follows: Sect.~\ref{sec:observations} gives details about the observations and the data reduction steps. We describe the analysis and the results in Sect.~\ref{sec:analysis}, including a characterization of the stellar activity from photometric and polarimetric time series (Sect.~\ref{sec:StellarActivityAnalysis}) and the analysis of the RV data (Sects.~\ref{sec:systematics} and \ref{sec:modelRV}). We follow with a discussion on the implications of those results in Sect.~\ref{sec:discussion} before concluding with Sect.~\ref{sec:conclusion}.

\section{Observations} \label{sec:observations}

\subsection{K2 photometry}\label{sec:K2data}

The \textit{Kepler} spacecraft observed \hbox{TRAPPIST-1} (K2-112) as part of Campaign 12 of its \textit{K2} mission \citep{howellK2MissionCharacterization2014}. This campaign lasted from December 15, 2016 to March 4, 2017, spanning a total of 78.9~d with a short gap lasting 5.3~d starting February 1, 2017 due to the spacecraft entering safe mode.

\begin{figure*}[ht]
\centering
\includegraphics[scale=0.5]{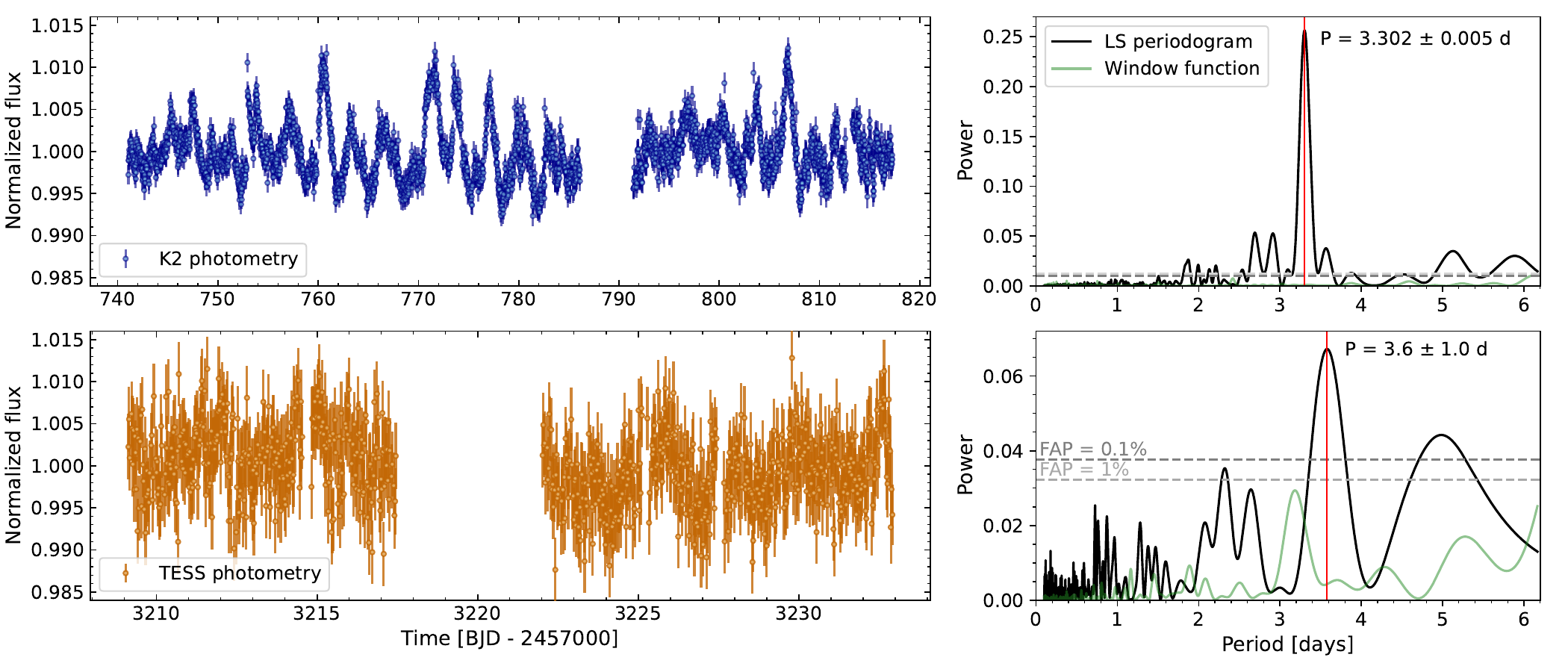}
\caption{Photometric data for TRAPPIST-1. Left panels: Normalized \textit{K2} \textit{(top)} and \textit{TESS} \textit{(bottom)} light curves. The \textit{TESS} light curve is binned to the 30-minutes cadence of \textit{K2}. Right panels: Lomb-Scargle periodograms with standard normalization (black) for the corresponding light curves. The periods associated with the maximum power (red lines) are identified at the top right, with uncertainties computed as described in Sect.~\ref{sec:PhotometryAnalysis}. The window functions (green) are shown as well as the 1\% (dashed, light gray) and 0.1\% (dashed, dark gray) false alarm probabilities (FAPs).}
\label{fig:photometry}
\end{figure*}

We used the long (30-minute) cadence light curve computed by \citet{lugerSevenplanetResonantChain2017}\footnote{\url{https://github.com/rodluger/trappist1}}, which is corrected for instrument systematics with the \texttt{EVEREST 2.0} pipeline \citep{lugerEVERESTPixelLevel2016, lugerUpdateEVERESTK22018}. We excluded the first 2.6~d of observation, which show a very steep increase in flux akin to instrument systematics, and corrected for a linear slope. The timestamps were converted to Barycentric Julian Day (BJD), and $3\sigma$ outliers were removed. Since the light curve is only used for stellar activity characterization, we removed all the in-transit points. Once normalized by the median flux, we obtained the light curve shown in the upper left panel of Fig.~\ref{fig:photometry}.

\subsection{\textit{TESS} photometry}\label{sec:TESSdata}

TRAPPIST-1 (TIC\,278892590, TOI-6838) was observed by \textit{TESS} in sector 70 from September 20 to October 15, 2023. Sector 70 covered the ecliptic as part of the second extended mission of \textit{TESS} and Year 6 of its overall mission. Using the \textit{TESS} Science Processing Operations Center \citep[SPOC;][]{jenkinsTESSScienceProcessing2016} data products, publicly available through the Mikulski Archive for Space Telescopes (MAST)\footnote{\url{http://archive.stsci.edu/tess/}}, we obtained the Presearch Data Conditioning Simple Aperture Photometry \citep[PDCSAP;][]{smithKeplerPresearchData2012, stumpeMultiscaleSystematicError2014} light curve sampled at 2-minute cadence. This light curve includes corrections for systematics seen across multiple quiet targets on the same detector channel \citep{kinemuchiDemystifyingKeplerData2012}. We verified that the stellar activity signal was not removed by the additional cleaning steps performed to obtain the PDCSAP light curve from the Simple Aperture Photometry (SAP). The light curve has a large 4.55-d gap due to the crossing of the Earth and the Moon, leaving us with $\sim$19~d of nearly-continuous data. We removed $3\sigma$ outliers, in-transit points (since we are only interested in stellar activity characterization) and binned to the same 30-minutes cadence as \textit{K2} to facilitate comparison. After normalization by the median flux, we got the light curve shown in the lower left panel of Fig.~\ref{fig:photometry}.

\subsection{SPIRou spectropolarimetry}\label{sec:RVdataSPIRou}

We obtained RV and polarimetric time series using SPIRou \citep{donatiSPIRouNIRVelocimetry2020}, a near-infrared high resolution (0.95 to 2.50\,$\mu$m, $R\sim70,000$) fiber-fed \'{e}chelle spectropolarimeter installed at the Cassegrain focus of the Canada-France-Hawai'i Telescope (CFHT) on Maunakea, Hawai'i. As part of the large program SPIRou Legacy Survey (SLS; ID~P42, PI: J.-F. Donati), of which the data are available on the Canadian Astronomy Data Centre (CADC) website\footnote{\url{https://www.cadc-ccda.hia-iha.nrc-cnrc.gc.ca}}, \hbox{TRAPPIST-1} was observed for 47 epochs between June 2019 and October 2021. Each epoch consists of a polarimetric sequence made up of four 674-s consecutive exposures with simultaneous Fabry-Pérot (FP) drift calibrations. During a polarimetric sequence, the positions of two rotating Fresnel rhombs are varied between exposures. With the two science fibers (A and B) receiving orthogonal polarization states, we gain access to both the circular polarization and the total intensity of the light beam (Stokes \textit{V} and \textit{I}).

Additionally, the dataset was complemented with 11 TRAPPIST-1\,b transit observations of 14 to 16 spectroscopic exposures (no polarization) of 513\,s each. These observations were made in the Dark mode, with the calibration channel set to Dark (instead of FP) to mitigate contamination of the science spectra from a prolonged contact with the relatively bright FP signal on the detector. All spectroscopy during transit events was obtained as part of PI programs led by X. Bonfils (ID~19AF07, 20BF18, 21BF16) and O. Lim (ID~20AC24, 20BC22).

Overall, the SPIRou dataset represents 364 spectra over 58 epochs. The observations were conducted with a median airmass of 1.18 (ranging from 1.1 to 2.3) and a median seeing of 0.56\,arcsec (ranging from 0.26 to 1.92\,arcsec), yielding a median S/N per pixel of 25 and 22 near 1.75\,$\mu$m for the polarimetric and spectroscopic exposures, respectively.

The SPIRou data were reduced with version 0.7.294 of A PipelinE to Reduce Observations (\texttt{APERO}\footnote{\url{https://github.com/njcuk9999/apero-drs}}), the standard pipeline for SPIRou data. We refer readers to \citet{cookAPEROPipelinEReduce2022} for details on the reduction procedure.

\subsubsection{SPIRou radial velocities}\label{sec:SPIRouRVs}

\begin{table*}[ht]
\caption{TRAPPIST-1 SPIRou and NIRPS RV observations.}
\label{table:RVobs}
\centering
\begin{tabular}{llccc}
\hline\hline
Instrument & Observation type & Epochs & Total exp. time [min]$^{(a)}$ & Median $\sigma_\mathrm{RV}$ [m\,s$^{-1}$] \\
\hline\noalign{\smallskip}
SPIRou & Polarimetric & 44 & 45 & 5.23 \\
& Spectroscopic (transit) & 11 & 120 & 2.90 \\\noalign{\smallskip}
NIRPS & Commissioning & 4 & 30 & 6.27 \\
& RV monitoring (SP1) & 24 & 15 & 10.76 \\
& Transit (SP3) & 16 & 115 & 4.23 \\ 
\hline
\end{tabular}
\tablefoot{SP1 and SP3 refer to the two sup-programs of the NIRPS-GTO described in Sect.~\ref{sec:NIRPSRV}.
$^{(a)}$Representative total exposure time for the associated observation type. Because of varying conditions (weather, technical issues, which planet is transiting, etc.), the exposure time can vary from one observation to another.}
\end{table*}

The RV time series was extracted from the telluric-corrected spectra using the line-by-line ({\tt LBL}, v0.65.003\footnote{\url{https://github.com/njcuk9999/lbl}}) method of \citet{artigauLinebylineVelocityMeasurements2022}. In this outlier-resistant framework, the Doppler shift is independently calculated for each spectral line using the \citet{bouchyFundamentalPhotonNoise2001} formalism, which is based on the difference between observations and a spectrum template of the star at high S/N. Because of the relative faintness of \hbox{TRAPPIST-1}, we instead used the closest star in spectral type to have a significant amount of high-S/N SPIRou observations (GJ~1002, an M5.5V; \citealt{walkerSpectroscopicSurvey1131983}), to construct the template. While GJ~1002 is slightly hotter than \hbox{TRAPPIST-1} \citepalias[$\Delta T_\mathrm{eff}\approx440$\,K;][]{agolRefiningTransittimingPhotometric2021, jahandarChemicalFingerprintsDwarfs2025}, it is the coldest star for which we have sufficient archival observations to construct the ``noise-free'' template required by the \citet{bouchyFundamentalPhotonNoise2001} formalism. The target was observed between July 2020 and July 2024 as part of the SLS and SPICE (SPIRou Legacy Survey -- Consolidation \& Enhancement; ID~P45, PI: J.-F. Donati) large programs, collecting a total of 861 polarimetric exposures of GJ~1002 over 216 epochs. We combined the individual telluric-corrected spectra from the \texttt{APERO} pipeline into a single high S/N template spectrum which was used to compute the \hbox{TRAPPIST-1} RVs within the {\tt LBL} framework. In practice, the slight spectral mismatch between the observations and the GJ~1002 template implies degraded RV precision. Still, we found GJ~1002 to yield more precise and accurate RVs than lower-S/N templates of later-type stars (including \hbox{TRAPPIST-1} itself).

\begin{figure*}[t]
    \centering
    \includegraphics[width=\linewidth]{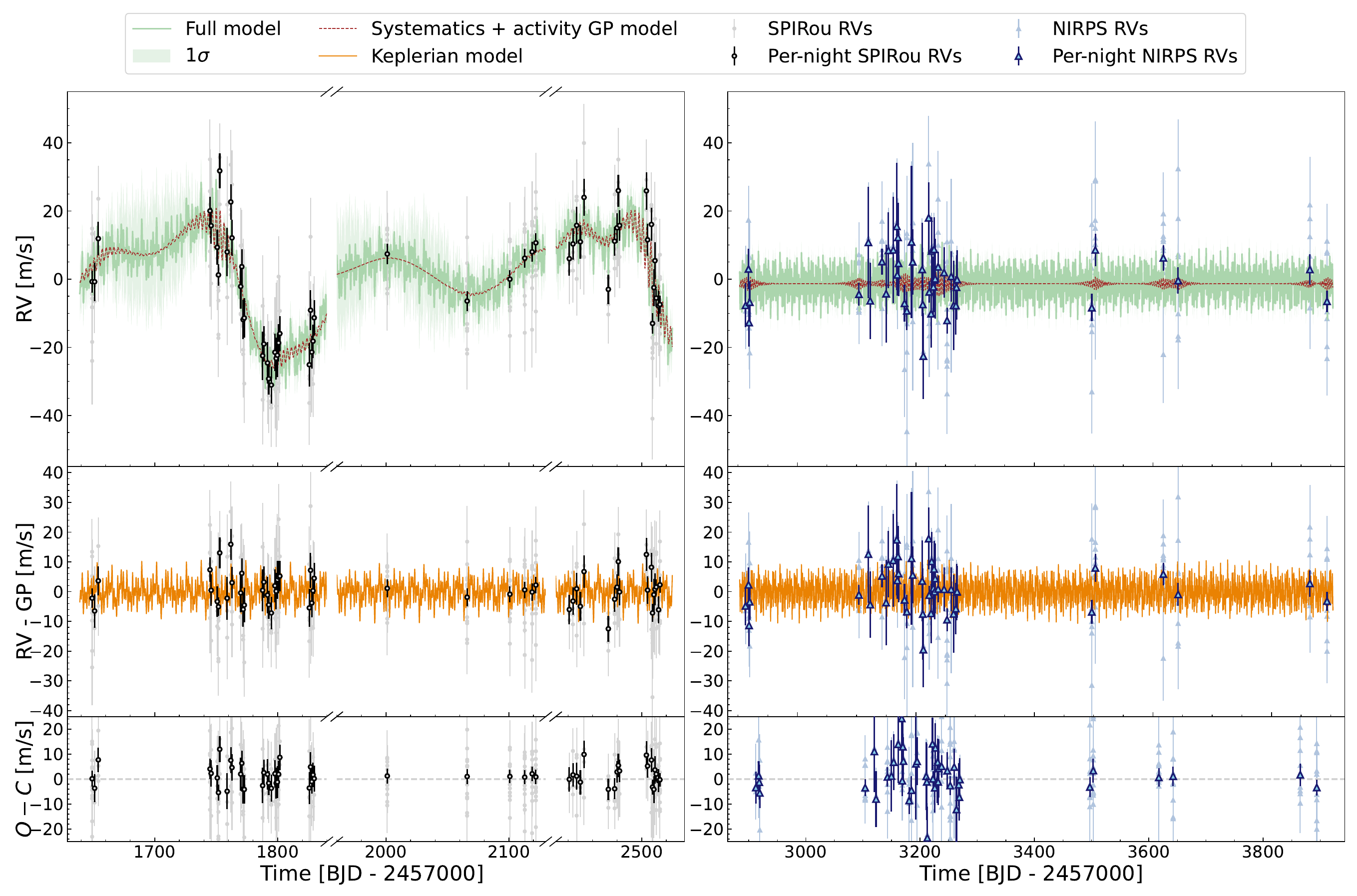}
    \caption{\textit{Top:} Full \hbox{TRAPPIST-1} SPIRou (left, circles) and NIRPS (right, triangles) RV time series, showing the individual exposures (pale transparent markers) and per-night bins (solid dark markers). The green line is the best-fit model (described in Sect.~\ref{sec:CombinedPlanetModel}), with its GP component modeling both systematics and stellar activity in maroon. \textit{Middle:} Same as the top panel but removing the GP, leaving only the planetary signal. The Keplerian component of the best-fit model is shown in orange. \textit{Bottom:} Residuals of the full model.}
    \label{fig:rv}
\end{figure*}

Out of the 364 TRAPPIST-1 spectra obtained with SPIRou, 349 passed the \texttt{APERO} quality check with 15 files failing the telluric correction step due to failed convergence of the telluric model. We calculated the {\tt LBL} RV measurements from the remaining spectra, applying a drift correction when the FP calibration is available. For the five epochs between December 8 and 12, 2019, we kept the non-drift corrected values as there were issues with the FP being exceptionally too bright during these dates. We additionally removed three outlying (3$\sigma$) RV points for the epoch of July 18, 2021. Overall, we were left with 346 RV points distributed in 55 epochs over $\sim2.4$~yr (see left panel of Fig.~\ref{fig:rv} and Table~\ref{tab:RVmeasurements}). The final RV measurements show a median precision of 10.90\,m\,s$^{-1}$ and a dispersion of 15.77\,m\,s$^{-1}$. When binning to one point per night, the median precision is 5.08\,m\,s$^{-1}$ and the dispersion is 15.96\,m\,s$^{-1}$. Specifically, the precision on the binned polarimetric exposures (total of $\sim$45\,min) is 5.23\,m\,s$^{-1}$, while the precision on the binned transits ($\sim$2\,h) is 2.90 m\,s$^{-1}$ (see Table~\ref{table:RVobs}).

\citet{bradyMeasuringObliquitiesTRAPPIST12023} measured the Rossiter-McLaughlin (RM) effect to have $\sim10$\,m\,s$^{-1}$ amplitude during the planetary transits of the \hbox{TRAPPIST-1} system. Considering the individual 513-s transit exposures and a $\sim$40-min transit duration for \hbox{TRAPPIST-1\,b}, we have four to five RV measurements that fall during the transit event for each of the 11 transit observations. Those measurements having a median uncertainty of 11.3\,m\,s$^{-1}$, we get $\sim8$\,m\,s$^{-1}$ precision on each side of the RM curve, insufficient to convincingly see the impact of the RM effect. Moreover, the RM signal is symmetric, with as much RV excess in the blue than the red (consistent with no spin-orbit misalignment, $\lambda=-2^{+17}_{-19}\,\degree$; \citealt{bradyMeasuringObliquitiesTRAPPIST12023}). We can thus bin the in-transit RV measurements without injecting a bias in our time series.

\subsection{NIRPS radial velocities}\label{sec:NIRPSRV}

Additional RV observations were obtained with the fiber-fed near-infrared high-resolution \'{e}chelle spectrograph NIRPS \citep{bouchyNIRPSJoiningHARPS2025}, in its high efficiency (HE) mode ($R\sim75,000$). Covering the $Y$, $J$, and $H$ bands (0.98 to 1.91\,$\mu$m) and equipped with adaptive optics, NIRPS is installed in parallel with HARPS \citep{mayorSettingNewStandards2003} at the European Southern Observatory (ESO) 3.6-m telescope at La Silla Observatory, Chile. \hbox{TRAPPIST-1} was observed during instrument commissioning (November and December 2022; PID: 60.A-9109(A)) and as part of the NIRPS Guaranteed Time Observation (NIRPS-GTO; PIs: F.~Bouchy and R.~Doyon) program, between June 2023 and August 2025. Observations were split between two of the NIRPS-GTO scientific sub-programs (SPs): SP1, a blind RV search for exoplanets around low-mass stars (PID: 111.251P.001, 112.25NZ.001), and SP3, an in-depth look into exoplanets through high-resolution time-series spectroscopy (PID: 111.2506.001, 112.25P3.001, 112.25P3.002, 112.25P3.003, 112.25P3.004) \citep{bouchyNIRPSJoiningHARPS2025, allartNIRPSDetectionDelayed2025}. 

SP1 observations of \hbox{TRAPPIST-1} were conducted as part of a survey studying the orbital architecture of systems around ultra-cool dwarfs (M7V and later), with the main goal of constraining the presence of massive planets well beyond the snow line. They consist of 28 epochs with 600 to 900\,s exposures (compared to $3\times600$\,s for commissioning). Within SP3, we observed three transits of TRAPPIST-1\,b, e and f, and four transits of TRAPPIST-1\,d and g. Each transit observation is made up of 12 (to 5) 600\,s (to 1800\,s) exposures, depending on the planet and the observing conditions. Overall, we have 188 spectra over 45 epochs with NIRPS. The observations were done with median airmass of 1.15 (ranging from 1.10 to 1.62) and median seeing of 1.26\,arcsec (ranging from 0.54 to 2.79\,arcsec), yielding a median S/N per pixel of 22 near 1.63\,$\mu$m. Each epoch has a corresponding HARPS observation which we did not include in our analysis due to extremely low S/N ($<2$ for the ten reddest orders in all exposures). 

The NIRPS spectra were reduced with the same data reduction software as the SPIRou ones \citep[\texttt{APERO} v0.7.294;][]{cookAPEROPipelinEReduce2022}. Out of the 188 spectra, 181 passed the \texttt{APERO} quality check with 7 files failing the telluric correction step due the telluric model not converging. The RVs were calculated from the remaining spectra, using the {\tt LBL} v0.65.003 algorithm \citep{artigauLinebylineVelocityMeasurements2022}. For the spectrum template, we used Proxima Centauri, the brightest M dwarf sample star to have both significant high-S/N NIRPS observations \citep{suarezmascarenoDivingPlanetarySystem2025} and the spectral type closest to that of \hbox{TRAPPIST-1} (M5.5 versus M8). We removed a $3\sigma$ outlier (epoch of December 1, 2022), leaving us with 180 RV measurements spread across 44 epochs over $\sim2.7$~yr (see right panel of Fig.~\ref{fig:rv} and Table~\ref{tab:RVmeasurements}). The final RVs show a median precision of 12.58\,m\,s$^{-1}$ and a dispersion of 12.41\,m\,s$^{-1}$. When binning the data to one point per night, the median precision is 8.73\,m\,s$^{-1}$ and the dispersion is 8.28\,m\,s$^{-1}$. Specifically, the transit data show 4.23\,m\,s$^{-1}$-precision for exposure times of $\sim$1.9\,h while the other observation have 10.08\,m\,s$^{-1}$-precision on 15-min median exposures (combining commissioning and SP1 data from Table~\ref{table:RVobs}). With even larger uncertainties than the SPIRou RV measurements, we conclude that the amplitude of the RM effect falls within errorbars for the individual planetary transits observed with NIRPS, justifying binning over the whole transit sequence. We can scale the NIRPS precision on 15-min exposures to reach the 45\,min total length of the SPIRou polarimetric sequences. Assuming three consecutive 15-min exposures, we would obtain a 5.82\,m\,s$^{-1}$ precision compared to 5.23\,m\,s$^{-1}$ for SPIRou thanks to its extended spectral coverage into the $K$ band.

\section{Analysis and results}\label{sec:analysis}

\subsection{Stellar activity}\label{sec:StellarActivityAnalysis}

TRAPPIST-1 exhibits significant levels of stellar activity, with recorded photometric and spectroscopic variability due to the presence of flares \citep{vidaFrequentFlaringTRAPPIST12017, limAtmosphericReconnaissanceTRAPPIST12023, howardCharacterizingNearinfraredSpectra2023, radicaPromisePerilStellar2025}, micro-flares \citep{berardoHubblesMultiyearSearch2026}, active regions on the surface of the star \citep{vidaFrequentFlaringTRAPPIST12017, morrisPossibleBrightStarspots2018} and likely condensate clouds
\citep[e.g.,][]{miles-paezTimeresolvedImagePolarimetry2019}. We characterized the stellar activity of \hbox{TRAPPIST-1} using various time series: long-term photometry and SPIRou spectropolarimetry with circular polarization.

\subsubsection{Stellar rotation period from photometry}\label{sec:PhotometryAnalysis}

The stellar rotation period $P_\mathrm{rot}$ is derived from the light curves obtained by \textit{K2} and \textit{TESS}, respectively providing the longest and most-recent quasi-continuous photometric time coverage of TRAPPIST-1. Previous studies independently derived $P_\mathrm{rot}$ from the \textit{K2} light curve using Discrete Fourier Transforms (DFT). While they obtained consistent values, their precision level differ significantly: $P_\mathrm{rot}=3.30\pm0.14$~d \citep{lugerSevenplanetResonantChain2017} and $3.295\pm0.003$~d \citep{vidaFrequentFlaringTRAPPIST12017}. With a generalized Lomb-Scargle periodogram (GLS;
\citealt{zechmeisterGeneralisedLombScarglePeriodogram2009}), \citet{diezalonsoCARMENESInputCatalogue2019} found a similar value of $P_\mathrm{rot}=3.304\pm0.011$~d.

To facilitate comparison between the \textit{K2} and \textit{TESS} light curves, we computed their rotation period using the same procedure. We employed GLS periodogram with standard normalization. Following \citet{vanderplasUnderstandingLombScarglePeriodogram2018}, $P_\mathrm{rot}$ is taken as the period associated with the highest power if it is above the 0.1\% false alarm probability (FAP) level and not associated with any aliases or peaks in the window function. Using an empirical method inspired by \citet{boyleStellarRotationStructure2023}, the uncertainty on $P_\mathrm{rot}$ is determined by computing many periodograms, masking for each a different 20\%-section of the light curve. To scan the whole light curve, we introduced a lag in the mask defined as the number of points corresponding to 20\% of $P_\mathrm{rot}$, looping back to the beginning once the end is reached to ensure all points have been masked an equal amount of times. Following this procedure, we obtained 116 period measurements for \textit{K2} and 34 for \textit{TESS} due mainly to their difference in time coverage. Taking their respective standard deviation yielded the uncertainties on the rotation periods.

The periodograms for the \textit{K2} and \textit{TESS} light curves are shown in the right panels of Fig.~\ref{fig:photometry} with the values and uncertainties recorded in Table~\ref{table:Prot}. For \textit{K2}, we obtain $P_{\text{rot}}=3.302\pm0.005$~d, in agreement with previous DFT computations. The precision is comparable to that of \citet{vidaFrequentFlaringTRAPPIST12017}. The period inferred from the \textit{TESS} light curve, while less precise due to lower S/N and shorter time coverage, is also in agreement ($P_{\text{rot}}=3.6\pm1.0$~d).

The photometric measurements of $P_\mathrm{rot}$ from \textit{K2} and \textit{TESS} are also in agreement with the recent value ($P_\mathrm{rot}=3.27\pm0.04$~d, see Table~\ref{table:Prot}) reported by \citet{berardoHubblesMultiyearSearch2026} based on a 5-yr monitoring campaign of the \hbox{TRAPPIST-1} Ly$\alpha$ flux with the Space Telescope Imaging Spectrograph on the \textit{Hubble} Space Telescope (HST/STIS). The presence of a $\sim3.3$-d periodicity in multiple time series spanning 2017 to 2023 strongly suggests the rotation of the star as its origin. Based on \textit{K2}, \textit{TESS} and HST/STIS, we get $3.3015\pm0.0050$~d from a weighted average, corresponding to the rotation period measured from \textit{K2} due to its much lower uncertainty compared to the other values. We adopt the flux-derived value $3.3015\pm0.0050$~d as reference for the following RV analysis.

\begin{table*}[ht]
\caption{TRAPPIST-1 stellar rotation periods.}
\label{table:Prot}
\centering
\begin{tabular}{lllcl}
\hline\hline
Dataset & Instrument & Method & Period [d] & Reference \\
\hline\noalign{\smallskip}
Photometry & \textit{K2} & DFT & $3.30\pm0.14$ & \citet{lugerSevenplanetResonantChain2017} \\
& & DFT & $3.295\pm0.003$ & \citet{vidaFrequentFlaringTRAPPIST12017} \\
& & GLS & $3.304\pm0.011$ & \citet{diezalonsoCARMENESInputCatalogue2019} \\
& & GLS & $3.302\pm0.005$ & This work \\
& \textit{TESS} & GLS & $3.6\pm1.0$ & This work \\ 
Ly$\alpha$ flux & HST/STIS & LS & $3.27\pm0.04$ & \citet{berardoHubblesMultiyearSearch2026} \\
RV & SPIRou+NIRPS & GP & $3.22^{+0.22}_{-0.20}$ & This work\\ \noalign{\smallskip}\hline
& & Weighted average$^{(a)}$ & $3.3015\pm0.0050$ & \\
\hline
\end{tabular}
\tablefoot{The methods used to compute the stellar rotation period are: discrete Fourier transforms (DFT), Gaussian processes (GP), generalized Lomb-Scargle (GLS) and Lomb-Scargle (LS) periodograms \citep{lombLeastsquaresFrequencyAnalysis1976, scargleStudiesAstronomicalTime1982}.
$^{(a)}$Computed using the periods derived in this work (\textit{K2}, \textit{TESS} and SPIRou+NIRPS) and from the Ly$\alpha$ time series.}
\end{table*}

\subsubsection{Upper limit on the large-scale magnetic field}\label{sec:polarimetry}

\begin{figure}[t]
    \centering
    \includegraphics[width=\linewidth]{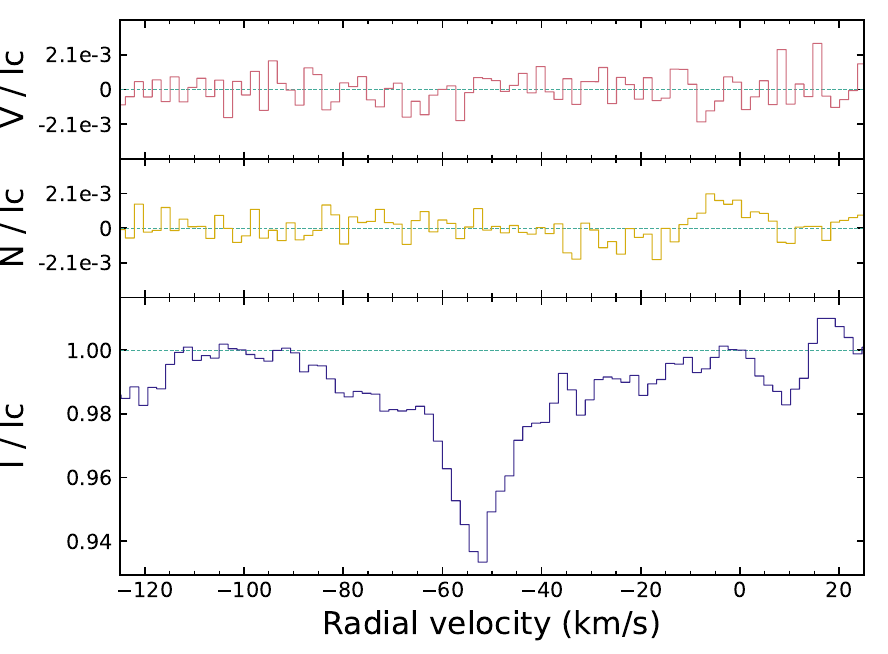}
    \caption{Average of all available LSD profiles (epochs 2019 and 2021). Circular polarization ($V/I_c$, top panel) does not show any signature at the RV of the intensity line profile ($I/I_c$, bottom panel). The middle panel shows the ``null'' profile \citep[$N/I_c$;][]{donatiSpectropolarimetricObservationsActive1997} supposed to feature noise only, unless defects (of instrumental or stellar origin) alter the polarimetric measurement. The non-detection places an upper limit on the longitudinal field $|B_l|<40$\,G ($3\sigma$), excluding an inclined dipolar field of polar strength $>0.1$\,kG.}
    \label{fig:lsd}
\end{figure}

Complete SPIRou polarimetric sequences were collected for \hbox{TRAPPIST-1} in 2019 (29 visits) and 2021 (13 visits). All circularly polarized spectra were processed with the least-square deconvolution method \citep[LSD;][]{donatiSpectropolarimetricObservationsActive1997, kochukhovLeastsquaresDeconvolutionStellar2010}, using a list of spectral lines delivered by the Vienna atomic line database \citep[VALD;][]{ryabchikovaMajorUpgradeVALD2015}, and computed from a MARCS atmospheric model with an effective temperature of 2500\,K and a surface gravity $\log g = 5.0$~cm~s$^{-2}$ \citep{gustafssonGridMARCSModel2008}, from which we kept atomic lines deeper than 5\% of the continuum level. In spite of the increase in S/N achieved through LSD, none of our observations led to a definite detection of a Zeeman signature, according to the $\chi^2$ statistical criterion of \cite{donatiSpectropolarimetricObservationsActive1997}. Longitudinal field ($B_l$) measurements \citep{reesLineFormationUnresolved1979,donatiSpectropolarimetricObservationsActive1997} were calculated on a velocity range going from --75 to --35 km\,s$^{-1}$. They provided us with an average value of --31\,G (after discarding four observations affected by a very low S/N), and a standard deviation of 45\,G over the 38 visits. The S/N increase obtained by stacking the available observations over rotational phase bins of various widths was  also not enough to reach the detection threshold. Finally, averaging all visits together leads again to a non-detection with a false alarm probability of 0.6 (Fig.~\ref{fig:lsd}), from which we derived an upper limit on the strength of the unsigned longitudinal field $|B_l| < 40$\,G ($3\sigma$). 

These upper limits allow us to exclude an inclined dipolar field with a polar strength greater than about 0.1\,kG. However, this field configuration is not the one most commonly reported for rapidly rotating, late-type M dwarfs. These stars usually have one of two types of magnetic field: a strong ($> 1$\,kG) axisymmetric dipolar field or a weaker multipolar and non-axisymmetric field \citep{morinLargescaleMagneticTopologies2010}. For TRAPPIST-1, we assume the stellar rotation axis to be perpendicular to our line of sight, since the planets' inclination is close to 90$\degree$ \citepalias{agolRefiningTransittimingPhotometric2021} and there is no evidence of spin-orbit misalignment in the system \citep[system obliquity consistent with zero;][]{hiranoEvidenceSpinOrbit2020, bradyMeasuringObliquitiesTRAPPIST12023}. Therefore, the detection of an axisymmetric dipole is very challenging: both poles are simultaneously visible, their polarimetric signatures overlapping and mostly canceling out. Still, since even a limited misalignment of the dipole and the rotation axis would lead to a detectable $B_l$, the case of a weaker multipolar field configuration is favored, similarly to the two stars of latest spectral type observed in spectropolarimetry prior to TRAPPIST-1 (vB~8 and vB~10, respectively an M7 and an M8; \citealt{morinLargescaleMagneticTopologies2010}). In this scenario, the sensitivity of our $B_l$ measurements may be too limited for a detection, according to the $B_l$ values reported by \cite{morinLargescaleMagneticTopologies2010} in this type of complex magnetic geometry. In summary, the upper limit obtained for \hbox{TRAPPIST-1} is compatible with available measurements for late M dwarfs with rotation periods of a few days. 

\subsection{SPIRou RV systematics}\label{sec:systematics}

\begin{figure*}
    \centering
    \includegraphics[width=\linewidth]{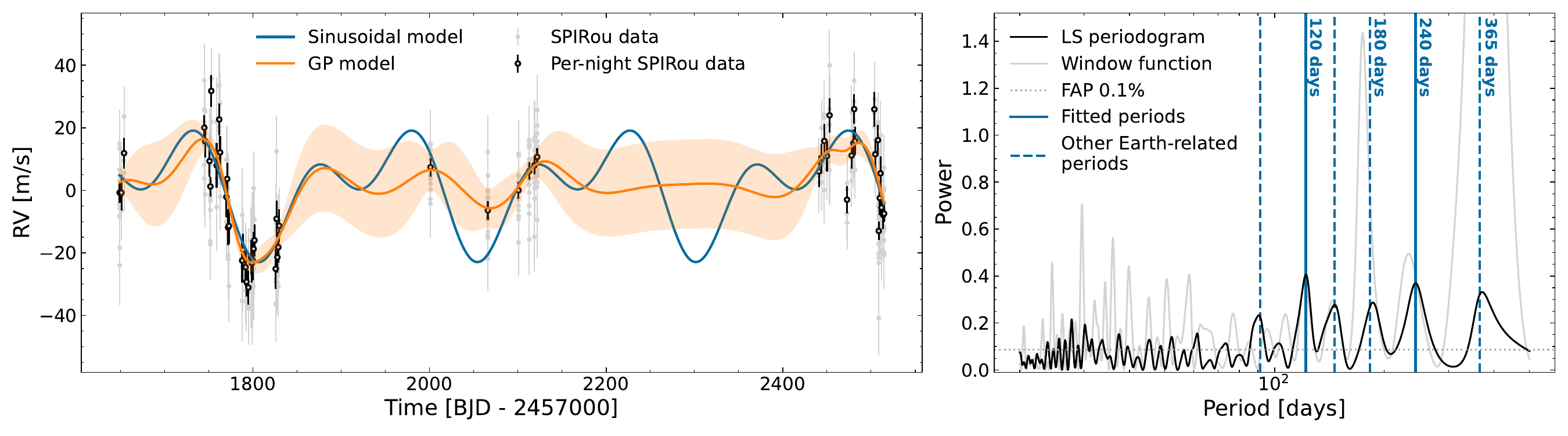}
    \caption{\textit{Left:} \hbox{TRAPPIST-1} SPIRou RV data overlaid with two models fitting for long-term systematics. The blue line is a sinusoidal model made up of the sum of two sinusoids (with periods $P_\mathrm{syst}$ and $P_\mathrm{syst}/2$, respectively) while the orange line represents a GP model with a stochastically-driven harmonic oscillator (SHO) kernel. \textit{Right:} GLS periodogram of the SPIRou RVs (black line) highlighting all the long-term periods attributed to the Earth's motion around the Sun (365-d harmonics; vertical blue lines) and the observation window function (gray line). The periods of the sinusoidal model are shown by the solid blue lines. Our baseline RV model uses the GP to fit the SPIRou systematics. Using the sinusoidal model instead is not favored based on difference in Bayesian evidence and does not affect the results (see Appendix~\ref{sec:AppendixRobustness}).}
    \label{fig:systematics}
\end{figure*}

Long-term variations are observed in the SPIRou RV measurements with, in particular, $\sim$40\,m\,s$^{-1}$ fluctuations near 2\,458\,800 and 2\,459\,500 BJD (see left panel of Fig.~\ref{fig:systematics}). The GLS periodogram reveals a number of significant periodicities on large timescales, many of which directly relate to the Earth's motion around the Sun. As shown in the right panel of Fig.~\ref{fig:systematics}, we find strong signal at 365\,d, its harmonics (notably $\sim$180, $\sim$120 and $\sim$90\,d), and at periods associated to the window function. Indeed, with SPIRou being mounted on the telescope only during bright time (near the full Moon), the regular monthly gaps in the time series lead to strong window function signals at the lunar cycle ($\sim30$\,d). 

We inferred that the strong long-term fluctuations seen in the \hbox{TRAPPIST-1} RV measurements are not astrophysical in origin by fitting a simple sinusoidal model to the data. The model is the sum of two sinusoids with their own amplitudes ($A$) and phases ($\Phi$), but with respective periods $P_\mathrm{syst}$ and $P_\mathrm{syst}/2$ (total of seven parameters after adding an instrumental mean $\phi$ and jitter $s$: $\{A_1,\,A_2,\,\Phi_1,\,\Phi_2,\,P_\mathrm{syst},\,\phi,\,s\}$). Optimizing the model with the nested sampling \citep{skillingNestedSampling2004, skillingNestedSamplingGeneral2006} algorithm \texttt{dynesty} \citep{speagleDYNESTYDynamicNested2020}, we find $P_\mathrm{syst}=247.34\pm1.40$\,d, meaning the two modes of the model ($P_\mathrm{syst}$ and $P_\mathrm{syst}/2$) correspond to window function and Earth aliases (solid blue lines, right panel of Fig.~\ref{fig:systematics}). This sinusoidal model reproduces the long-term RV trend well (left panel of Fig.~\ref{fig:systematics}). We thus conclude that they are systematics resulting from an imperfect telluric correction. 

TRAPPIST-1 is particularly faint (even in the NIR with $H=10.7$\,mag; \citealt{skrutskieTwoMicronAll2006}) and therefore affected by persistence effects (seen in SPIRou for $H>10$\,mag; \citealt{donatiSPIRouNIRVelocimetry2020}). Persistence is the phenomenon of having the trace spectrum of a previously-observed star still present on the infrared detector, diluting the spectral lines of the currently-observed star \citep[e.g.,][]{barilCharacterizationPersistenceWIRCams2008, artigauOpticalNearinfraredRadial2018}. With time, the trace spectrum gradually fades away, but the asymmetric persistence signature on the detector complicates data processing steps such as telluric correction, leading to telluric residuals being left in the spectra (see Appendix~\ref{sec:persistence} for more details on persistence). For faint stars, those residuals can have a significant impact on the extracted RV measurements, as evidenced by the systematics in the \hbox{TRAPPIST-1} time series (Fig.~\ref{fig:systematics}).

To model systematics, we also explored other methods such as Gaussian Processes \citep[GPs;][]{rasmussenGaussianProcessesMachine2006, ambikasaranFastDirectMethods2015}, which have the flexibility to model complex structures without the drawback of fitting for the large number of parameters required by a combination of sinusoidal signals. The GP is described by $\mathcal{GP}(\phi,\,\text{\textbf{K}}_{ij})$, with a mean function $\phi$, which we take to be a constant, and a covariance matrix \textbf{K}$_{ij}$ \citep{hwangHowUseGP2023}. The covariance matrix is defined by 
\begin{equation}\label{eq:covMatrixSyst}
    \text{\textbf{K}}_{ij} = k_\mathrm{syst}(t_i,\,t_j) + \delta_{ij}(\sigma(t_i)^2 + s^2)
\end{equation}
having the systematics covariance kernel specified by $k_\mathrm{syst}(t_i,\,t_j)$. A Kronecker delta term $\delta_{ij}$ adds the measurement uncertainties $\sigma(t_i)$ and an jitter $s$ to the diagonal elements of \text{\textbf{K}}$_{ij}$, which represent the variance of each point with itself. For the SPIRou systematics, we used the stochastically-driven harmonic oscillator (SHO) kernel implemented in \texttt{celerite2} \citep[\texttt{SHOTerm};][]{foreman-mackeyFastScalableGaussian2017, foreman-mackeyScalableBackpropagationGaussian2018}. A typical SHO kernel is described by three hyperparameters: $\sigma$, the standard deviation of the process, $\tau$, the damping timescale of the process, and $\rho$, the undamped period of the oscillator. We optimized for \{$\sigma_\mathrm{syst}$, $\tau_\mathrm{syst}$, $\rho_\mathrm{syst}$, $\phi$, $s$\} with the same nested sampling routine as the sinusoidal model. We used log-uniform priors on $\sigma_\mathrm{syst}$, $\tau_\mathrm{syst}$ and $\rho_\mathrm{syst}$ ($\mathcal{LU}(1,\,50)$\,m\,s$^{-1}$, $\mathcal{LU}(50,\,400)$\,d and $\mathcal{LU}(20,\,400)$\,d, respectively), ensuring that the minimum allowed $\tau_\mathrm{syst}$ is larger than the period of the outermost planet \citepalias[TRAPPIST-1\,h, $\sim18.8$~d;][]{agolRefiningTransittimingPhotometric2021}. The best-fit GP model has $\rho_\mathrm{syst}\sim120$~d and reproduces the long-term RV trends, similarly to the sinusoidal model (left panel of Fig.~\ref{fig:systematics}). For our baseline RV model of the TRAPPIST-1 system, we selected to model SPIRou systematics with a GP component. Alternative treatment with the sinusoidal model is explored in Appendix~\ref{sec:AppendixRobustness}, yielding the same results.

As per \citet{bouchyNIRPSJoiningHARPS2025}, we know that persistence is also present on the NIRPS detector. However, we see no evidence for systematics resulting from telluric residuals in the RV measurements obtained with this instrument. One possible explanation is that we observe less persistence in NIRPS than SPIRou (see Appendix~\ref{sec:persistence}), leading to a more robust telluric correction. Still, we cannot exclude the possibility that some lower-amplitude systematics are present in the NIRPS RVs and we simply do not have the precision required to isolate them.

\subsection{Modeling the RVs}\label{sec:modelRV}

The RVs of the \hbox{TRAPPIST-1} system include three components: systematics, stellar activity and planetary signals. In our baseline model, the RV variations caused by instrument systematics (as seen in Sect.~\ref{sec:systematics}) and stellar activity are modeled with specific covariance kernels applied to a single GP described in Sect.~\ref{sec:GPmodel}. For the planetary component, Sects.~\ref{sec:SevenKepPlanetModel} and \ref{sec:CombinedPlanetModel} explore different ways to efficiently model the RV signal created by the seven \hbox{TRAPPIST-1} planets.

\subsubsection{GP model}\label{sec:GPmodel}

In our baseline model, a GP model is used to model the effects of both systematics and stellar activity on the RV data. In Sect.~\ref{sec:systematics}, we described the systematics seen in SPIRou with a GP characterized by a SHO kernel, $k_\mathrm{syst}$. To model stellar activity, we added a second SHO kernel \citep[from \texttt{celerite2};][]{foreman-mackeyFastScalableGaussian2017, foreman-mackeyScalableBackpropagationGaussian2018}, $k_\mathrm{act}$, to the GP covariance matrix described in equation~\ref{eq:covMatrixSyst}. Each RV dataset is described by its own GP, $\mathcal{GP}_\mathrm{SPIRou}(\phi_\mathrm{SPIRou},\,\mathbf{K}_{\mathrm{SPIRou},\,ij})$ and $\mathcal{GP}_\mathrm{NIRPS}(\phi_\mathrm{NIRPS},\,\mathbf{K}_{\mathrm{NIRPS},\,ij})$, with their covariance matrices comprising instrument-specific elements (systematics and jitter) and a common stellar activity kernel $k_\mathrm{act}$:
\begin{align}\label{eq:covMatrixfull}
    &\mathbf{K}_{\mathrm{SPIRou},\,ij} =k_\mathrm{syst}(t_{i},\,t_{j}) + k_\mathrm{act}(t_i,\,t_j)  \nonumber \\
    &\;\;\;\;\;\;\;\;\;\;\;\;\;\;\;\;\;+\delta_{ij}(\sigma_\mathrm{SPIRou}(t_i)^2 + s_\mathrm{SPIRou}^2) \\
    &\mathbf{K}_{\mathrm{NIRPS},\,ij} =k_\mathrm{act}(t_i,\,t_j) + \delta_{ij}(\sigma_\mathrm{NIRPS}(t_i)^2 + s_\mathrm{NIRPS}^2) \nonumber
\end{align}
The stellar activity kernel is described by a standard deviation $\sigma_\mathrm{act}$, interpreted as the amplitude of the stellar activity-induced RV signal, a damping timescale $\tau_\mathrm{act}$, indicating the evolution timescale for structures (spots, faculae) on the stellar surface, and an activity period $\rho_\mathrm{act}$, seen as the rotation period of the star. Including the instrument-specific parameters like the respective constant mean functions $\phi$ and jitters $s$, as well as the SPIRou systematics (Sect.~\ref{sec:systematics}), the full GP model is described by the parameters \{$\ln\sigma_\mathrm{syst}$, $\ln\tau_\mathrm{syst}$, $\ln\rho_\mathrm{syst}$, $\ln\sigma_\mathrm{act}$, $\ln\tau_\mathrm{act}$, $\rho_\mathrm{act}$, $\phi_\mathrm{SPIRou}$, $\phi_\mathrm{NIRPS}$, $\ln{s_\mathrm{SPIRou}}$, $\ln{s_\mathrm{NIRPS}}$\}. Note that the same $\sigma_\mathrm{act}$ is used for SPIRou and NIRPS, as they both operate in the NIR.

We explored an alternative treatment of stellar activity in Appendix \ref{sec:AppendixRobustness}, using a GP with  quasi-periodic (QP) kernel instead of SHO \citep{foreman-mackeyFastScalableGaussian2017}.  We find equivalent results and differences in Bayesian evidence favoring our baseline model with a SHO for stellar activity.

RV analyses typically use activity indicators to inform their GP model and avoid them fitting planetary signals. Activity indicators notably include photometric light curves \citep{aigrainSimpleMethodEstimate2012} and other time series derived directly from the available spectra: H$\alpha$, chromatic RV index and differential line width \citep[CRX and dLW;][]{zechmeisterSpectrumRadialVelocity2018}, differential temperature ($d$Temp; \citealt{artigauMeasuringSubKelvinVariations2024}), $\log R'_\mathrm{HK}$ \citep{hartmannAnalysisVaughanPrestonSurvey1984, noyesRotationConvectionMagnetic1984, suarezmascarenoRotationPeriodsLatetype2015}, etc. The indicators can serve as ``training'' datasets for the GP \citep[e.g.,][]{cloutierCharacterizationK218Multiplanetary2017, cloutierConfirmationRadialVelocity2019, cadieuxDetailedArchitecture98592025}, where the GP is first fitted on the indicators to inform priors for the RV fit. Alternatively, the GP model can be simultaneously fitted on the indicators and the RV, sharing some physical parameters such as timescale and rotation period  \citep[e.g.,][]{bauerCARMENESSearchExoplanets2020, lacedelliTransitingRockySuperEarth2025}, or through a multi-dimensional framework \citep[e.g.,][]{gonzalezhernandezSubEarthmassPlanetOrbiting2024, suarezmascarenoDivingPlanetarySystem2025, stefanovHADESRVProgramme2025} as described by \citet{rajpaulGaussianProcessFramework2015}. 

However, we were not successful in retrieving reliable spectral activity indicators from the \hbox{TRAPPIST-1} spectra. Because of the persistence effect \citep{barilCharacterizationPersistenceWIRCams2008, artigauH4RGCharacterizationHighresolution2018} seen in both SPIRou and NIRPS (Sect.~\ref{sec:systematics}, Appendix~\ref{sec:persistence}), particularly affecting faint targets, the spectral lines are diluted to varying levels from one observation to the next. Persistence affects all spectroscopically-derived activity indicators, as they rely on measuring changes in the shape of the lines. Correcting for persistence exceeds the scope of this paper, so we decided to conservatively exclude all spectroscopic indicators from the analysis.

When trying to fit the \hbox{TRAPPIST-1} light curves with a GP model, we found \textit{TESS} to be too noisy to allow convergence and \textit{K2} to be too detailed \citep[stellar rotation, micro-variability;][]{morrisPossibleBrightStarspots2018} to reliably model with a simple SHO kernel. To retain physically-interpretable hyperparameters, the \textit{K2} light curve (upper left panel, Fig.~\ref{fig:photometry}) would require a more complex model which would then be fine-tuned to a specific activity pattern predating the first RV observation by more than two years. Given a spot evolution timescale comparable to the stellar rotation \citep{morrisPossibleBrightStarspots2018}, basing the RV activity GP on the \textit{K2} light curve is ill-advised, especially considering the limited amount and sparseness of the RV observations (99 epochs over $\sim6$\,yr). 

We used the flux-derived stellar rotation period ($P_\mathrm{rot}=3.3015\pm0.0050$~d; Sect.~\ref{sec:PhotometryAnalysis}) to inform the period of the GP activity kernel $\rho_\mathrm{act}$. We put a relatively loose constraint on $\rho_\mathrm{act}$ with a normal prior of $\mathcal{N}(3.3,\,0.5)$~d. For the logarithmic evolution timescale $\ln\tau_\mathrm{act}$, we used a uniform prior with a lower bound at the stellar rotation period, equivalent to $\mathcal{LU}(3.3,\,10^3)$\,days. All remaining parameters ($\ln\sigma_\mathrm{act}$, $\phi_\mathrm{SPIRou}$, $\phi_\mathrm{NIRPS}$, $\ln s_\mathrm{SPIRou}$ and $\ln s_\mathrm{NIRPS}$) have uniform priors as listed in Table~\ref{table:parameters}.

\subsubsection{Seven Keplerian planetary model}\label{sec:SevenKepPlanetModel}

\begin{figure*}[ht]
\centering
\includegraphics[width=\linewidth]{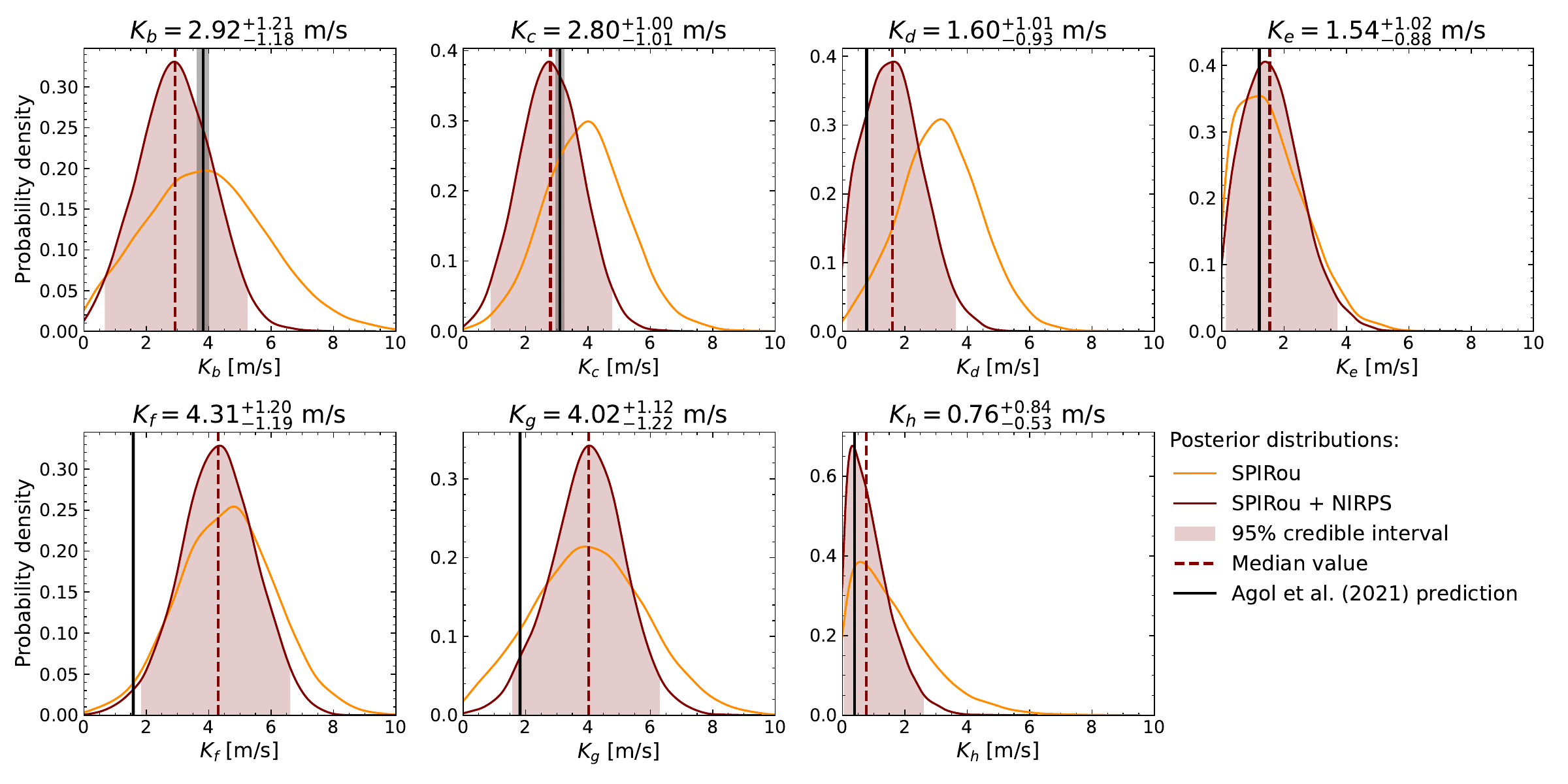}
\caption{Marginalized posterior distributions for the RV semi-amplitude of each \hbox{TRAPPIST-1} planet. We show the posterior distributions for the SPIRou-only (orange) and SPIRou+NIRPS (dark red) datasets. For the SPIRou+NIRPS case, we highlight the 95\% credible intervals (shaded region) and the median values (dashed line). Even if none of the planets are reliably detected to $>3\sigma$, the presumptive distributions are consistent with the predictions from the \citetalias{agolRefiningTransittimingPhotometric2021} TTV model (black lines with a gray shaded region for the uncertainty, which is too small to see except for $K_{\mathrm{b}}$ and $K_{\mathrm{c}}$). The semi-amplitudes of TRAPPIST-1\,f and g are overestimated, which can be attributed to statistical fluctuations and covariance with the RV semi-amplitude of other planets.}
\label{fig:indPlanets}
\end{figure*}

The TTVs observed in the \hbox{TRAPPIST-1} system \citepalias{agolRefiningTransittimingPhotometric2021} indicate that the planets do not evolve on independent Keplerian orbits \citep{agolTransitTimingDurationVariations2018}. Still, when comparing the stellar RV predicted by the \hbox{TRAPPIST-1} N-body simulation of \citetalias{agolRefiningTransittimingPhotometric2021} to a circular multi-Keplerian model, we found excellent agreement between the two: over a simulation of 100~d, we obtained a standard deviation of $7.38\times10^{-3}$\,m\,s$^{-1}$ for the residuals. Even over a longer timescale (1600~d), where the longitude of periapsis of each planet precesses, the standard deviation of the residuals only increased to $\sim2.7\times10^{-2}$\,m\,s$^{-1}$. Considering our best RV precision of $\sim5$\,m\,s$^{-1}$ (Table~\ref{table:RVobs}), we decided to adopt the circular multi-Keplerian approximation as our planetary model.

The planetary model is the sum of seven Keplerian models, one for each planet with index $j$. We assumed circular orbits (eccentricity $e_j=0$ and argument of periapsis arbitrarily set to $\omega_j=\pi/2$\;rad) for all the \hbox{TRAPPIST-1} planets, motivated by the comparison with the N-body simulation and the small eccentricities derived from stability analyses and tidal evolution models of the \hbox{TRAPPIST-1} system \citep[$e<0.01$;][]{gillonSevenTemperateTerrestrial2017, lugerSevenplanetResonantChain2017, turbetModelingClimateDiversity2018}. We took the orbital periods $P_j$ and times of inferior conjunction $t_{0,\,j}$ from the linear transit timings simulation of \citet{agolUpdatedForecastTRAPPIST12024}, which represent the mean ephemeris. We decided to fix the values of $P_j$ and $t_{0,\,j}$, considering the $<0.001\%$ precision on $P_j$ \citep{gillonSevenTemperateTerrestrial2017, lugerSevenplanetResonantChain2017} and the $\sim10^{-2}$\,m\,s$^{-1}$ deviation caused by the use of Keplerians with periodic transit timings $t_{0,\,j}$ instead of an N-body simulation. We also checked that fitting $t_{0,\,j}$ with normal prior distributions taken from \citet{agolUpdatedForecastTRAPPIST12024} does not change the results. Therefore, we only optimized the planetary model for the RV semi-amplitude $K_j$ of each planet \{$K_\mathrm{b}$, $K_\mathrm{c}$, $K_\mathrm{d}$, $K_\mathrm{e}$, $K_\mathrm{f}$, $K_\mathrm{g}$, $K_\mathrm{h}$\}, using wide and uniform priors $\mathcal{U}(0,\,10)$\,m\,s$^{-1}$ within the dynamic nested sampling framework of \texttt{dynesty} to improve posterior density estimation \citep{higsonDynamicNestedSampling2019, speagleDYNESTYDynamicNested2020}. The baseline GP model described in Sect.~\ref{sec:GPmodel} is fitted conjointly to the seven Keplerian planetary model on two datasets: SPIRou only and SPIRou+NIRPS (nightly binned). We found that fitting only the NIRPS RV measurements does not constrain the model due to the median precision being much lower (8.73\,m\,s$^{-1}$ vs 5.08\,m\,s$^{-1}$ for SPIRou) for this dataset.

The full posterior results of the complete model fitting the seven Keplerian (``individual'') planetary signal on the SPIRou+NIRPS dataset are listed in Table~\ref{table:parameters}. We find that our SPIRou and NIRPS RV data do not contain enough information to individually detect the signals of the \hbox{TRAPPIST-1} planets. For each, we find consistent results between the model fitted on SPIRou-only and SPIRou+NIRPS RV measurements (Fig.~\ref{fig:indPlanets}), but none of the planets are reliably detected to more than $3\sigma$ (from 1.1 to 2.8$\sigma$, excluding \hbox{TRAPPIST-1\,f and g}). Still, the presumptive distributions of all the planets are consistent within their 95\% credible interval with the RV semi-amplitudes predicted by the TTV model of \citetalias{agolRefiningTransittimingPhotometric2021}. The only exceptions are \hbox{TRAPPIST-1}\,f and~g, for which we measured $K_{\mathrm{f}}=4.3\pm1.2$\,m\,s$^{-1}$ and $K_\mathrm{g}=4.0^{+1.1}_{-1.2}$\,m\,s$^{-1}$ compared to the predicted $1.581\pm0.047$\,m\,s$^{-1}$ and $1.822\pm0.052$\,m\,s$^{-1}$, respectively.

We consider the possibility that $K_\mathrm{f}$ and $K_\mathrm{g}$ are overestimated. From statistical fluctuations alone, we expect 1 out of 20 independent parameters to fall outside the 95\% credible interval. In this case, the parameters are not independent: a positive covariance is notably observed between $K_\mathrm{f}$ and $K_\mathrm{g}$. To assess the impact of the covariances, we performed an injection-recovery test to measure the frequency of measuring a RV semi-amplitude 2$\sigma$ away from its injected value based on our data. To do so, we simulated \hbox{TRAPPIST-1} RV measurements with the same timestamps and uncertainties as our SPIRou and NIRPS observations. For the underlying model, we took the best-fit GP component of the ``individual planets'' model (see Table~\ref{table:parameters}) and added Keplerian planetary signals with RV semi-amplitudes based on the \citetalias{agolRefiningTransittimingPhotometric2021} predictions. Assuming a Gaussian noise distribution, we created a 100 datasets of noisy RVs from the measured uncertainties. We then fitted the model described above to each individual dataset and found that 23/100 have one $K_j$ differing from the injected RV semi-amplitude by 2$\sigma$. An additional 5/100 have two differing $K_j$, totaling some overestimation 28\% of the time. In most of the datasets, $K_\mathrm{f}$ is the culprit, being overestimated 11 times out of 100, almost twice as much as $K_\mathrm{g}$, the next worst-performing $K_j$. Thus, it is not improbable to measure with our dataset a different $K_\mathrm{f}$ and $K_\mathrm{g}$ from the \citetalias{agolRefiningTransittimingPhotometric2021} predictions. Considering statistical fluctuations and interplay with the signals from other planets, from stellar activity, or the RV time sampling itself, we conclude that there are, with this first analysis, no significant evidence suggesting that the \hbox{TRAPPIST-1} planets' masses differ from the ones inferred with TTV modeling. 

To estimate the number of RV measurements required to significantly detect the individual TRAPPIST-1 planets ($3\sigma$), we used the formalism of \citet[][equation 18 therein]{cloutierQuantifyingObservationalEffort2018}. For each planet, we computed the effective RV uncertainty for a representative measurement of our dataset, based on the level of significance obtained with the seven Keplerian planetary model described in this section (see $K_j$ values of Fig.~\ref{fig:indPlanets} and Table~\ref{table:parameters}). Assuming the RV semi-amplitudes predicted by \citetalias{agolRefiningTransittimingPhotometric2021}, we found that 150 RV measurements of similar quality would allow the detection of TRAPPIST-1\,b, but that $\sim1000$ would be needed to reach the smallest signal of the system, TRAPPIST-1\,h.

\begin{table*}[h!]
\caption{Results of the RV model of TRAPPIST-1, fitting for the individual (Sect.~\ref{sec:SevenKepPlanetModel}) or the combined (Sect.~\ref{sec:CombinedPlanetModel}) signal from all the planets on the SPIRou+NIRPS dataset.}
\label{table:parameters}
\centering
\begin{tabular}{llcrr}
\hline\hline
Parameter & Description & Prior & \multicolumn{2}{c}{Posterior} \\
\hline\noalign{\smallskip}
\textit{GP model} & & & \textit{Individual} & \textit{Combined} \\[.02cm]
    --- \textit{Systematics} & & & & \\
    $\ln\sigma_\mathrm{syst}$ & Log-standard deviation of systematics [$\ln$ m\,s$^{-1}$] 
        & $\mathcal{U}$(0, 4)
        & $2.73^{+0.31}_{-0.25}$
        & $2.74^{+0.30}_{-0.24}$ \\ [.1cm]
    $\ln\tau_\mathrm{syst}$ & log-timescale of systematics [$\ln{\rm d}$]
        & $\mathcal{U}$(3, 6)
        & $3.85^{+1.07}_{-0.61}$
        & $3.82^{+1.08}_{-0.60}$ \\ [.1cm]
    $\ln\rho_\mathrm{syst}$ & log-period of systematics [$\ln{\rm d}$]
        & $\mathcal{U}$(4, 6)
        & $4.83^{+0.32}_{-0.30}$
        & $4.82^{+0.40}_{-0.36}$ \\ [.15cm]
    --- \textit{Stellar activity} & & & & \\
    $\ln\sigma_\mathrm{act}$ & Log-standard deviation of activity [$\ln$ m\,s$^{-1}$] 
        & $\mathcal{U}$(0, 4)
        & $1.10^{+0.34}_{-0.55}$
        & $1.21^{+0.32}_{-0.63}$ \\ [.1cm]
    $\ln\tau_\mathrm{act}$ & log-timescale of activity [$\ln{\rm d}$]
        & $\mathcal{U}$(1.2, 6.76)
        & $3.12^{+2.01}_{-1.24}$
        & $2.71^{+2.02}_{-0.98}$ \\ [.1cm]
    $\rho_\mathrm{act}$ & Period of activity [d]
        & $\mathcal{N}$(3.3, 0.5)
        & $3.09^{+0.30}_{-0.12}$
        & $3.22^{+0.22}_{-0.20}$ \\ [.15cm]
\hline\noalign{\smallskip}
    \textit{Planetary model} & & & & \\[.02cm]
    For $j$ in \{b, c, d, e, f, g, h\}: & & & & \\ [.1cm]
    ~~~~$P_j$ & Orbital period [d] 
        & (a) & ... & ... \\ [.1cm]
    ~~~~$t_{0,\,j}$ & Time of inferior conjunction [BJD]
        & (a) & ... & ... \\ [.1cm]
    ~~~~$e_j$ & Eccentricity
        & 0 & ... & ... \\ [.1cm]
    ~~~~$\omega_j$ & Argument of periapsis [rad] 
        & $\pi/2$ & ... & ... \\ [.1cm]
    ~~~~$K_j/K_{\mathrm{b}}$ & RV semi-amplitude of planet $j$ relative to TRAPPIST-1\,b$^{(c)}$
        & (b) & ... & ... \\ [.15cm]
    $K_{\mathrm{b}}$ & RV semi-amplitude of TRAPPIST-1\,b [m\,s$^{-1}$]
        & $\mathcal{U}$(0, 10) 
        & $2.92^{+1.21}_{-1.18}$
        & $3.65^{+0.78}_{-0.83}$ \\ [.1cm]
    $K_{\mathrm{c}}$ & RV semi-amplitude of TRAPPIST-1\,c [m\,s$^{-1}$]
        & $\mathcal{U}(0, 10)$
        & $2.80^{+1.00}_{-1.01}$ 
        & ... \\ [.1cm]
    $K_{\mathrm{d}}$ & RV semi-amplitude of TRAPPIST-1\,d [m\,s$^{-1}$]
        & $\mathcal{U}$(0, 10)
        & $1.60^{+1.01}_{-0.93}$
        & ... \\ [.1cm]
    $K_{\mathrm{e}}$ & RV semi-amplitude of TRAPPIST-1\,e [m\,s$^{-1}$]
        & $\mathcal{U}$(0, 10)
        & $1.54^{+1.02}_{-0.88}$
        & ... \\ [.1cm]
    $K_{\mathrm{f}}$ & RV semi-amplitude of TRAPPIST-1\,f [m\,s$^{-1}$]
        & $\mathcal{U}$(0, 10)
        & $4.31^{+1.20}_{-1.19}$
        & ... \\ [.1cm]
    $K_{\mathrm{g}}$ & RV semi-amplitude of TRAPPIST-1\,g [m\,s$^{-1}$]
        & $\mathcal{U}$(0, 10)
        & $4.02^{+1.12}_{-1.22}$
        & ... \\ [.1cm]
    $K_{\mathrm{h}}$ & RV semi-amplitude of TRAPPIST-1\,h [m\,s$^{-1}$]
        & $\mathcal{U}$(0, 10)
        & $0.76^{+0.84}_{-0.53}$
        & ... \\ [.15cm]
\hline\noalign{\smallskip}
    \textit{Instrumental parameters} & & & & \\ [.02cm]
    $\phi_\mathrm{SPIRou}$ & Mean SPIRou RV deviation [m\,s$^{-1}$]
        & $\mathcal{U}$(--20, 20)
        & $1.52^{+4.72}_{-4.41}$
        & $2.19^{+4.82}_{-4.54}$ \\ [.1cm]
    $\phi_\mathrm{NIRPS}$ & Mean NIRPS RV deviation [m\,s$^{-1}$]
        & $\mathcal{U}$(--20, 20)
        & $-0.83^{+1.00}_{-1.04}$ 
        & $-1.33^{+1.14}_{-1.08}$ \\ [.1cm]
    $\ln{s_\mathrm{SPIRou}}$ & Log-jitter for SPIRou  [$\ln$ m\,s$^{-1}$] 
        & $\mathcal{U}$(--9.2, 2.3)
        & $-0.01^{+1.63}_{-6.11}$
        & $0.22^{+1.43}_{-6.39}$ \\ [.1cm]
    $\ln{s_\mathrm{NIRPS}}$ & Log-jitter for NIRPS  [$\ln$ m\,s$^{-1}$] 
        & $\mathcal{U}$(--9.2, 2.3)
        & $-4.29^{+3.41}_{-3.37}$
        & $-4.66^{+3.65}_{-3.16}$ \\ [.1cm]
\hline
\end{tabular}
\tablefoot{$\mathcal{U}$(min, max) refers to a uniform prior between values min and max and $\mathcal{N}(\mu, \sigma)$, to a normal distribution of mean $\mu$ and standard deviation $\sigma$. The listed posteriors values refer to the median of the marginalized distributions along with their uncertainties from the 16th and 84th percentiles.
(a) Taken from the linear transit timing simulation of \citet{agolUpdatedForecastTRAPPIST12024} 
(b) Based on the predicted RV semi-amplitude of \citet{agolRefiningTransittimingPhotometric2021} 
$^{(c)}$Only used for the combined model}
\end{table*}

\subsubsection{Combined planetary model}\label{sec:CombinedPlanetModel}

As a second step, we looked for the combined RV planetary signal of the \hbox{TRAPPIST-1} system, assuming again circular orbits. While the signal from the individual planets is predicted to range from 0.4 to 3.8\,m\,s$^{-1}$, they sum to an amplitude of 5--10\,m\,s$^{-1}$ \citepalias{agolRefiningTransittimingPhotometric2021}, within reach of our RV precision. 

\begin{figure}[t!]
\centering
\includegraphics[width=\linewidth]{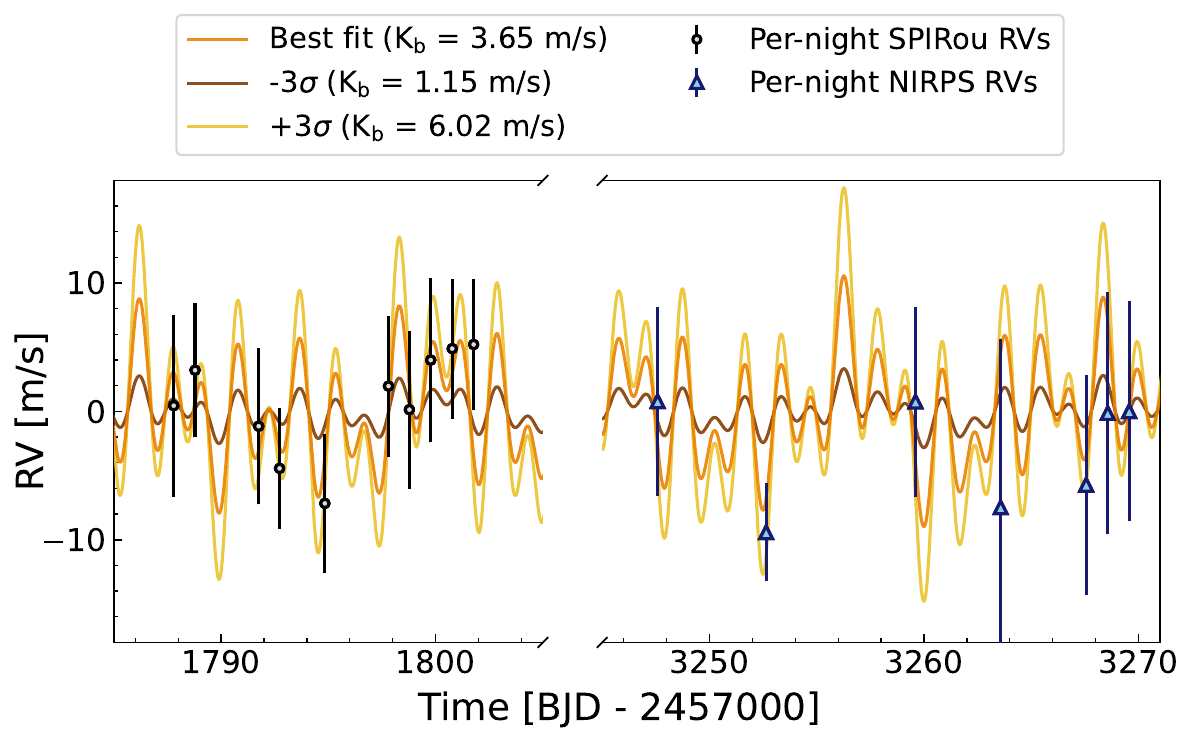}
\caption{Relation between the total RV planetary signal and the RV semi-amplitude of TRAPPIST-1\,b, $K_\mathrm{b}$, in the combined planetary model of Sect.~\ref{sec:CombinedPlanetModel}. Using the relative planetary masses provided by TTVs, we construct a model for the RV signal of the whole system, fitting its absolute scale. We show the model for the best-fit $K_\mathrm{b}$ (orange line, equivalent to the Keplerian model of Fig.~\ref{fig:rv}) and its $3\sigma$ upper (yellow line) and lower (brown line) values. A small sub-sample of per-night SPIRou (black circles) and NIRPS (blue triangles) RV measurements are shown as reference.}
\label{fig:demoCombinedModel}
\end{figure}

TTV models have highly degenerate parameter spaces, allowing multiple possible combinations of planetary masses. However, all solutions follow a linear relationship under a tight mass ratio between planets, as recently demonstrated by \citet{jinIllusoryPrecisionTTV2026} for the Kepler--9 system, confirming the theoretical prediction of \citet{lithwickExtractingPlanetMass2012}. We can thus confidently rely of the relative masses provided by TTV modeling, using RV to scale this combined signal to the absolute planetary mass values. To search for the combined RV signal of all the \hbox{TRAPPIST-1} planets, we employed an alternative parametrization, redefining $K_j$ using the relative masses derived from TTVs:
\begin{equation}\label{eq:ratio}
    K_j\longrightarrow K_\mathrm{b} \cdot (K_{j,\,\mathrm{TTV}}/K_{\mathrm{b,\,TTV}})
\end{equation}
where $K_{j,\,\mathrm{TTV}}$ is the RV semi-amplitude predicted from the TTV mass of planet $j$ \citepalias[see Table~10 of][]{agolRefiningTransittimingPhotometric2021}. This reparameterization simplifies the planetary model, reducing from seven parameters ($K_j$, $j=\{\mathrm{b},\,\mathrm{c},\,\mathrm{d},\,\mathrm{e},\,\mathrm{f},\,\mathrm{g},\,\mathrm{h}\}$) to only one, the RV semi-amplitude of TRAPPIST-1\,b \{$K_\mathrm{b}$\}. Note that here, $K_\mathrm{b}$ is a proxy for the whole planetary system, as illustrated in Fig.~\ref{fig:demoCombinedModel}.  

We adopted a wide and uniform prior for $K_{\mathrm{b}}$, $\mathcal{U}(0,\,10)$\,m\,s$^{-1}$, and optimized the combined planetary model jointly to the baseline GP model described in Sect.~\ref{sec:GPmodel} (including SPIRou systematics and stellar activity), using the dynamic nested sampling framework of \texttt{dynesty} \citep{speagleDYNESTYDynamicNested2020}. As presented in Appendix~\ref{sec:AppendixRobustness}, we confirmed the robustness of the results to treatments of systematics and stellar activity differing from the baseline model. Again, the model is first fitted to the (nightly-binned) SPIRou RVs before adding the NIRPS RVs for a complete SPIRou+NIRPS fit, excluding a NIRPS-only fit as we found again that the dataset is not constraining on its own. As shown by the posterior distributions of Fig.~\ref{fig:rvcorner}, we found perfectly consistent results between the two datasets. Therefore, any result cited below is from the SPIRou+NIRPS fit unless stated otherwise.

By employing a modeling framework combining the signal of the seven planets, we successfully detect the planetary signal of the \hbox{TRAPPIST-1} system (Fig.~\ref{fig:rvcorner}). We find $K_\mathrm{b}=3.65^{+0.78}_{-0.83}$\,m\,s$^{-1}$, in agreement with the TTV prediction of $3.82\pm0.19$\,m\,s$^{-1}$ \citepalias{agolRefiningTransittimingPhotometric2021}. We note that the signal is detected from the SPIRou data alone ($K_\mathrm{b}=4.86\pm1.26$\,m\,s$^{-1}$), but that adding the NIRPS RVs increases both the significance and accuracy of the detection. For the SPIRou systematics model, we find an amplitude of $\sim$15\,m\,s$^{-1}$ ($\ln\sigma_\mathrm{syst}=2.74^{+0.30}_{-0.24}$), a period of $\sim$120~d ($\ln\rho_\mathrm{syst}=4.82^{+0.40}_{-0.36}$) and a timescale of $\sim$50~d ($\ln\tau_\mathrm{syst}=3.82^{+1.08}_{-0.60}$), confirming that known planets and stellar activity, with their much shorter periods, are not fitted by this GP component. 

For the stellar activity, we recover a period of $\rho_\mathrm{act}=3.22^{+0.22}_{-0.20}$~d, in agreement with the rotation period derived in Sect.~\ref{sec:PhotometryAnalysis} from flux time series ($P_\mathrm{rot}=3.3015\pm0.0050$\,d). While the $\rho_\mathrm{act}$ posterior is already more constrained than the informed prior used in the analysis ($\mathcal{N}(3.3,\,0.5)$\,d), we confirmed the detection of the stellar rotation period from the RV measurements alone by obtaining the same result using instead the uniform prior $\mathcal{U}(2,\,10)$\,d. We recover an upper bound on the amplitude of stellar activity ($\sim6.9$\,m\,s$^{-1}$ ($3\sigma$), $\ln\sigma_\mathrm{act}=1.21^{+0.32}_{-0.63}$), which includes the predicted $\sim$5\,m\,s$^{-1}$ from the synthesized RV activity curve of \citet{kleinSimulatingRadialVelocity2019}. We obtain a stellar activity timescale of $\sim$15~d ($\ln\tau_\mathrm{act}=2.71^{+2.02}_{-0.98}$), corresponding to activity patterns remaining coherent over $\sim$4--5 rotations on the surface of TRAPPIST-1. Such a short activity timescale is consistent with very few patterns repeating in the \textit{K2} light curve, as observed by \citet{morrisPossibleBrightStarspots2018}. However, it is significantly shorter than the decay time of strongly magnetic stars such as M0--M3 slow rotators observed with SPIRou \citep{donatiMagneticFieldsRotation2023}, including Gl~388 (AD~Leo), a M3V star with similar rotation period ($P_\mathrm{rot}\sim2.2$~d; \citealt{carmonaNearIROpticalRadial2023} and references therein) as TRAPPIST-1. The short activity timescale measured thus suggests that TRAPPIST-1 might not be strongly magnetic, in accordance with the weak multipolar configuration favored by the analysis of the large-scale field (Sect.~\ref{sec:polarimetry}).

\begin{figure}[t]
\centering
\includegraphics[width=\linewidth]{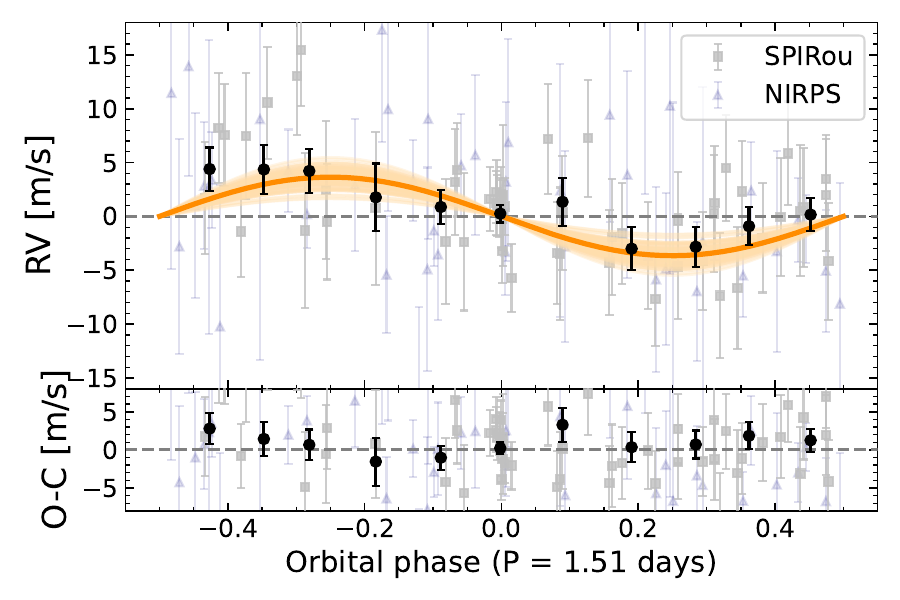}
\caption{Radial velocity of TRAPPIST-1\,b, in m\,s$^{-1}$, as a function of the orbital phase of the planet. The gray squares (SPIRou) and purple triangles (NIRPS) are the per-night RV points, after removing the GP and the other planets' signal obtained from the best-fit model. By binning the RVs, we get the black points. The Keplerian shown in orange is computed with the best-fit value of $K_{\mathrm{b}}$, while the light orange lines represent models randomly drawn from its posterior distribution.}
\label{fig:rvfolded}
\end{figure}

The posterior values of the combined planetary model fitted on the SPIRou+NIRPS dataset are recorded in Table~\ref{table:parameters}, including the prior used for each parameter. The best-fit model is overlaid to the RV time series in Fig.~\ref{fig:rv}. For visibility purposes, we isolated the signal of TRAPPIST-1\,b, subtracting the GP model and the other planets' Keplerians from the RV data. After phase-folding onto the orbital period of TRAPPIST-1\,b and binning the data, the recovered planetary signal becomes more apparent (Fig.~\ref{fig:rvfolded}).

\subsubsection{Model comparison}\label{sec:modelComparision}

\begin{figure*}[ht]
\sidecaption
\centering
\includegraphics[width=12cm]{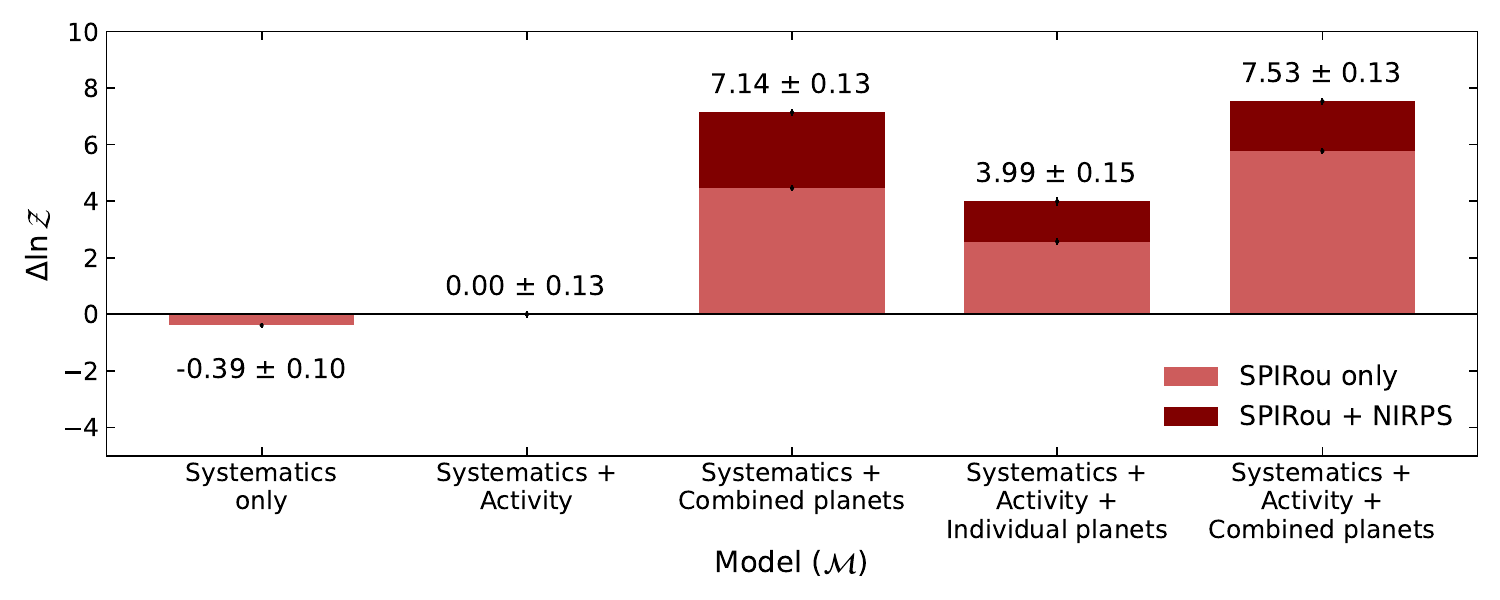}
\caption{Difference in Bayesian log-evidence ($\Delta\ln\mathcal{Z}$) between various RV models $\mathcal{M}$ and one including only the systematics and stellar activity components. A larger positive (negative) $\Delta\ln\mathcal{Z}$ indicates a more favored (disfavored) model $\mathcal{M}$. The results for the SPIRou RV measurements are in light red, while the dark red shows the results from the SPIRou + NIRPS dataset. A black errorbar indicates the uncertainties for each measurement, which are too small to see. All the models including a planetary component are favored. Based on the thresholds of \citet{trottaBayesSkyBayesian2008}, we have moderate evidence for the presence of planets if we consider them individually ($\Delta\ln\mathcal{Z}=3.99\pm0.15$, 54:1 odds), which increases to a strong evidence by considering their combined RV signature ($\Delta\ln\mathcal{Z}=7.53\pm0.13$, 1860:1 odds).}
\label{fig:evidenceComparison}
\end{figure*}

In the interest of evaluating the significance of the planetary detection, we compared the Bayesian log-evidences of the models from Sects.~\ref{sec:SevenKepPlanetModel} and~\ref{sec:CombinedPlanetModel}. Those two models include the systematics, stellar activity and planetary components, fitting respectively individual and combined planetary signals. We also fitted different levels of model complexity, including: only systematics; systematics + stellar activity (no planets); and systematics + ``combined planets'' (no stellar activity), following the parametrization of equation~\ref{eq:ratio}. To compare the models, we computed the difference in log-evidence relative to the model without a planetary component (for a model $\mathcal{M}$, $\Delta\ln\mathcal{Z}=\ln\mathcal{Z}_\mathcal{M} - \ln\mathcal{Z}_\mathrm{syst+act}$). The results are shown in Fig.~\ref{fig:evidenceComparison} for the SPIRou-only and SPIRou+NIRPS datasets.

Comparing the $\Delta\ln\mathcal{Z}$ presented in Fig.~\ref{fig:evidenceComparison}, it is clear that the models including all the components (systematics, stellar activity and planetary signals) are favored. Based on the thresholds defined by \citet{trottaBayesSkyBayesian2008}, the evidence for the presence of planets in the \hbox{TRAPPIST-1} system is moderate when they are modeled individually ($\Delta\ln\mathcal{Z}=3.99\pm0.15>2.5$, 54:1 odds) as in Sect~\ref{sec:SevenKepPlanetModel}, but increases to strong when their combined RV signature is considered ($\Delta\ln\mathcal{Z}=7.53\pm0.13>5$, 1860:1 odds), following Sect.~\ref{sec:CombinedPlanetModel}. For both models, adding the NIRPS RVs increases the detection significance, but we note that the $\Delta\ln\mathcal{Z}>5$ threshold is reached from the SPIRou data alone for the ``combined planets'' model (Sect.~\ref{sec:CombinedPlanetModel}). On the other hand, despite a strong detection of the activity period $\rho_\mathrm{act}$, we found inconclusive evidence for stellar activity in our RV time series.  Comparing the $\Delta\ln\mathcal{Z}$ of the full model with combined planets to the model without activity, we obtained $\Delta\ln\mathcal{Z}=0.39\pm0.18$ (3:2 odds). Still, knowing from photometry that \hbox{TRAPPIST-1} is an active star (Sect.~\ref{sec:PhotometryAnalysis}) and seeing a slight improvement in Bayesian evidence by including stellar activity in the RV model (Fig.~\ref{fig:evidenceComparison}), we selected the three-component model (systematics + stellar activity + ``combined planets'', described in Sect.~\ref{sec:CombinedPlanetModel}) as our best-fit model. We validated that including or not a stellar activity component does not affect the results of the systematics and the planetary components.

\section{Discussion}\label{sec:discussion}

\subsection{Long-term monitoring of the stellar rotation}\label{sec:discussProt}

The rotation period of \hbox{TRAPPIST-1} has long been debated, the main disagreement being between the light curves obtained by \textit{Spitzer} \citep{gillonSevenTemperateTerrestrial2017, delrezEarly2017Observations2018} and \textit{K2} \citep{lugerSevenplanetResonantChain2017}. Indeed, the clear 3.3-d period inferred from \textit{K2} \citep[$P_{K2}$;][]{lugerSevenplanetResonantChain2017, vidaFrequentFlaringTRAPPIST12017, dmitrienkoActivityM8Dwarf2018, diezalonsoCARMENESInputCatalogue2019} is not found in \textit{Spitzer}, with \citet{roettenbacherStellarActivityTRAPPIST12017} reporting $0.819\pm0.015$~d ($\sim$\,1/4 $P_{K2}$), while \citet{delrezEarly2017Observations2018} and \citet{morrisPossibleBrightStarspots2018} find no significant periodicity. These results have been explained by an evolving stellar surface \citep{roettenbacherStellarActivityTRAPPIST12017} and the presence of a few bright stellar spots \citep{morrisPossibleBrightStarspots2018}. Moreover, the flares observed in the \textit{K2} light curve have been associated to moments of increased flux baseline, suggesting that the 3.3-d signal could be the characteristic timescale of active regions, capturing their growth and decay, rather than the stellar rotation \citep{morrisPossibleBrightStarspots2018}. To disentangle the nature of the periodic signals detected from TRAPPIST-1, long-term monitoring of the star is required.

Long-term monitoring of TRAPPIST-1 was achieved by \citet{berardoHubblesMultiyearSearch2026}, who looked at the Ly$\alpha$ flux with HST/STIS and found $P_\mathrm{rot}=3.27\pm0.04$~d. We are able to further confirm the stability of the stellar signal from our SPIRou+NIRPS RV measurements: our best-fit RV model (Sect.~\ref{sec:CombinedPlanetModel}) yields a precise stellar activity period of $\rho_\mathrm{act}=3.22^{+0.22}_{-0.20}$\,d. The remarkable stability of the detected period over more than six~years across all time series strongly supports a stellar-rotation origin \citep{morrisPossibleBrightStarspots2018}. We also note that while photometry is sensitive to heterogeneities covering the whole stellar disk, RV is mostly sensitive to the equatorial region due to it having more significant displacement in our line of sight. As all the different time series yield coherent periods, we conclude that there is currently no sign of differential rotation for TRAPPIST-1. Overall, we adopt the weighted average of the periods derived in this paper (\textit{K2}, \textit{TESS} and SPIRou+NIRPS) and from the Ly$\alpha$ time series \citep{berardoHubblesMultiyearSearch2026} as the rotation period of TRAPPIST-1, $P_\mathrm{rot}=3.3015\pm0.0050$~d (Table~\ref{table:Prot}). 

To further validate our RV-derived rotation period, we computed the stellar rotational velocity ($v\sin i$) assuming the rotation axis of the star to be perpendicular to our line of sight ($\sin i\approx1$) and \hbox{R$_{\star}=0.1192\pm0.0013$\,R$_{\odot}$} \citepalias{agolRefiningTransittimingPhotometric2021}. We obtained \hbox{$v\sin i = 1.88\pm0.13$\,km\,s$^{-1}$}, in agreement with the value measured from the Rossiter–McLaughlin (RM) effect of the system with MAROON-X \citep[$v\sin i = 2.1\pm0.3$ km\,s$^{-1}$ ;][]{bradyMeasuringObliquitiesTRAPPIST12023} and SPIRou+IRD \citep[$v\sin i = 1.81^{+0.40}_{-0.33}$ km\,s$^{-1}$ ;][]{LimThesis2026}. We note that \citet{bradyMeasuringObliquitiesTRAPPIST12023} included in their model the differential rotation velocity parameter $\alpha$. While they found an $\alpha$ consistent with zero, meaning they were unable to detect differential rotation for TRAPPIST-1, a model with $\alpha=0$ (and separate obliquities for the planets) instead resulted in $v\sin i=2.5\pm0.4$\,km\,s$^{-1}$, $1.5\sigma$ away from our $v\sin i$ derived with the RV stellar rotation period.

Stellar rotational velocity can also be directly measured from line profile broadening. In this case, other processes such as instrumental effects (e.g., persistence), stellar temperature, turbulence and limb-darkening must be disentangled from rotational broadening \citep{reinersCARMENESSearchExoplanets2018, varasCARMENESSearchExoplanets2026}. Typically, the relation between $v\sin i$ and the width of the spectral lines is determined with a reference template: a star of the same spectral type, assumed to have $v\sin i\sim 0$\,km\,s$^{-1}$ \citep{houdebineObservationModellingMainsequence2010}. Obtaining such a template is especially challenging for late-type M dwarfs like TRAPPIST-1. Since non-rotating M8V stellar templates are not readily available for SPIRou or NIRPS, we determined that the direct spectroscopic measurement of the TRAPPIST-1 $v\sin i$ would require a dedicated analysis beyond the scope of this paper.

\subsection{RV detection of the \hbox{TRAPPIST-1} planetary system}

We recover the combined signature of the whole \hbox{TRAPPIST-1} planetary system in RV with our best-fit model of Sect.~\ref{sec:CombinedPlanetModel}. In terms of RV semi-amplitude, we find $K_{\mathrm{b}}=3.65^{+0.78}_{-0.83}$\,m\,s$^{-1}$ from the SPIRou+NIRPS dataset, corresponding to a planetary mass of $1.31\pm0.29$\,M$_\oplus$ for TRAPPIST-1\,b. This value is a proxy for the whole planetary system, the semi-amplitudes $K_j$ (and thus planetary masses $M_{p,\,j}$) of all the other planets being recovered through equation~\ref{eq:ratio} (see Table~\ref{tab:KandMp}). The mass obtained from the SPIRou and NIRPS RV measurements is in agreement with the TTV value of $M_{p,\,\mathrm{b}}=1.374\pm0.069$\,M$_\oplus$ \citepalias{agolRefiningTransittimingPhotometric2021}.

\begin{table}[t]
\caption{RV semi-amplitudes and masses measured for the TRAPPIST-1 planets.}
\label{tab:KandMp}
\centering
\begin{tabular}{lccc}
\hline\hline
Planet & $K_j$ [m\,s$^{-1}$] & $M_\text{RV}$ [M$_\oplus$] & $M_\text{TTV}$ [M$_\oplus$]$^{(a)}$ \\
\hline\noalign{\smallskip}
b & $3.65^{+0.78}_{-0.83}$ & $1.31\pm0.29$ & $1.374\pm0.069$ \\ [0.13cm]
c & $2.97^{+0.63}_{-0.67}$ & $1.25\pm0.28$ & $1.308\pm0.056$ \\ [0.13cm]
d & $0.74\pm0.17$ & $0.37\pm0.08$ & $0.388\pm0.012$ \\ [0.13cm]
e & $1.15\pm0.26$ & $0.66\pm0.15$ & $0.692\pm0.022$ \\ [0.13cm]
f & $1.51\pm0.33$ & $0.99\pm0.22$ & $1.039\pm0.031$ \\ [0.13cm]
g & $1.74\pm0.38$ & $1.26\pm0.28$ & $1.321\pm0.038$ \\ [0.13cm]
h & $0.37\pm0.08$ & $0.31\pm0.07$ & $0.326\pm0.020$ \\
\hline
\end{tabular}
\tablefoot{The $K_j$ and $M_\text{RV}$ values are listed for the best-fit RV model, described in Sect.~\ref{sec:CombinedPlanetModel}. These values depend on the relative planetary masses obtained from TTV, being 
derived from a single constraint on $K_{\mathrm{b}}$ using equation~\ref{eq:ratio}. For constraints on the RV semi-amplitudes without TTV-derived assumptions, refer to Table~\ref{table:parameters}.
$^{(a)}$Table 6 of \citetalias{agolRefiningTransittimingPhotometric2021}}
\end{table}

It should be noted that the RV model allowing the detection of the TRAPPIST-1 system (described in Sect.~\ref{sec:CombinedPlanetModel}) is not independent from TTVs. Indeed, to construct the complete planetary signal, we enforced that the planets' relative masses follow that obtained from the TTV analysis \citepalias{agolRefiningTransittimingPhotometric2021}. This choice was made following an initial analysis which revealed that the individual RV semi-amplitudes $K_j$ are in relatively good agreement with predictions based on the TTV model (Sect.~\ref{sec:SevenKepPlanetModel}). Recent results from \citet{jinIllusoryPrecisionTTV2026} also indicate that the TTV planetary mass ratios are reliable. Thus, the RV detection is a first estimation of what a complete TTV+RV fit would yield. While such model exceeds the scope of this paper, we show with our analysis that high precision RV observations for the \hbox{TRAPPIST-1} system are consistent with the TTV solution of \citetalias{agolRefiningTransittimingPhotometric2021}.

Beyond validating the planetary masses, RVs provide highly complementary information to systems with TTVs. As highlighted by \citet{almenaraAbsoluteMassesRadii2015}, detecting dynamical interactions allows the determination of key parameters, such as $M_p/M_\star$, from photometry alone. Obtaining the absolute masses of the objects in the system then requires a measurement of the stellar mass, which can be obtained from stellar evolution models (e.g., \citealt{vangrootelStellarParametersTrappist12018}) or empirical relations (e.g., \citealt{mannHowConstrainYour2019}). Alternatively, RV can be used to directly measure the system scale, allowing the determination of the absolute stellar mass $M_\star$ \citep{agolDetectingTerrestrialPlanets2005,Montet2013, almenaraAbsoluteMassesRadii2015}. The expression for the RV semi-amplitude $K$ is given by \citep[][equation 2.27]{perrymanExoplanetHandbook2018}
\begin{equation}
    K = \left(\frac{2\pi G}{P}\right)^{1/3}\frac{M_p\sin i}{(M_\star+M_p)^{2/3}}\frac{1}{(1-e^2)^{1/2}},
\end{equation}
where $G$ is the gravitational constant, $P$ the orbital period, $M_p$ the planet mass, $i$ the orbital inclination and $e$ the eccentricity. Considering that $K$ is measured through RV and that we get the mass ratio $q=M_p/M_\star$ from TTVs, we can rewrite the stellar mass as
\begin{equation}
    M_\star=\left(\frac{P}{2\pi G}\right)\frac{(1-e^2)^{3/2}}{\sin^3i}\frac{(1+q)^2}{q^3}K^3.
\end{equation}
We inputted the parameters for TRAPPIST-1\,b: $P_b$ from the linear ephemerids of \citet{agolUpdatedForecastTRAPPIST12024}, $e_b=0$ (see Sect.~\ref{sec:SevenKepPlanetModel}), $i_b=89.728\pm0.165\degree$, $q_b=(4.596\pm0.198)\times10^{-5}$ \citepalias{agolRefiningTransittimingPhotometric2021} and $K_{\mathrm{b}}=3.65^{+0.78}_{-0.83}$\,m\,s$^{-1}$. We obtain the absolute stellar mass $M_\star=0.078^{+0.064}_{-0.042}$\,M$_\odot$ for TRAPPIST-1, in agreement with model-based determinations ($0.089\pm0.006$\,M$_\odot$; \citealt{vangrootelStellarParametersTrappist12018}) and values derived from empirical relations \citepalias[$0.0898\pm0.0023$\,M$_\odot$;][]{agolRefiningTransittimingPhotometric2021}. While the precision is far from comparable due to the limited RV precision \citep[e.g.,][]{almenaraAbsoluteMassesRadii2015}, this absolute stellar mass is free of any of the assumptions that affect the stellar evolution models in particular. It also provides an important comparison point for the lower-mass end of the \citet{mannHowConstrainYour2019} empirical relation, which covers down to 0.075\,M$_\odot$ and relies on the characterization of stars in stellar binaries. Since single stars might evolve differently than binary stars \citep{marchantEvolutionBinaryStars2025}, having model-free mass determination of single stars like TRAPPIST-1 is highly valuable.

\subsection{Constraints on the presence of giant planets with NIRPS}\label{sec:detectionOtherPlanets}

The RV monitoring of \hbox{TRAPPIST-1} enables the characterization of the architecture of its planetary system. Notably, we can constrain the presence of massive planets beyond the snow line. These distant planets are challenging to observe through transit due to their limited transit observation windows or possible non-transiting alignments. While TTVs can be used to infer the presence of additional planets, they are biased towards near-resonant orbits and specific dynamical architecture. Indeed, the presence of generalized three-body Laplace relations between adjacent triplets of planets (GLRs, \citealt{papaloizouThreeBodyResonances2015}) was used to predict the orbital period of TRAPPIST-1\,h before its confirmation by \citet{lugerSevenplanetResonantChain2017}. An eighth planet has also been hypothesized due to the presence of GLRs among the seven known planets \citep{pletserExponentialDistanceRelation2017, kippingPredictingOrbitTRAPPIST1i2018}, but targeted searches of the TTVs at the suggested resonant periods have not yielded any conclusive evidence so far \citepalias{agolRefiningTransittimingPhotometric2021}. Because of the size of the parameter space in TTV dynamical modeling, no exhaustive search has been conducted. With our RV data, we can place meaningful constraints on the presence of additional planets in the \hbox{TRAPPIST-1} system without being limited by inclination or mean motion resonance geometries.

We decided against using the SPIRou RVs to search for additional planets in the \hbox{TRAPPIST-1} system. The GP component used to model the SPIRou long-term systematics significantly hinders our ability to blindly detect signals on long timescales. With a more stable RV time series, NIRPS allows one to constrain the presence of massive planets at large orbital distances. 

To determine the detection sensitivity of the NIRPS RV data, we performed a set of injection-recovery tests. To inject the Keplerian signals, we followed a procedure inspired by \citet{gonzalezhernandezSubEarthmassPlanetOrbiting2024}: we first constructed an activity-only RV time series by taking the GP component of the best-fit RV model (combining the planetary signals, see Sect.~\ref{sec:CombinedPlanetModel}) and adding white noise based on the standard deviation of its residuals. Then, following \citet{cloutierMorePreciseMass2021} and \citet{cherubimTOI1695WaterWorld2023}, we injected a single Keplerian signal into $10^5$ realizations of the noisy RVs. We sampled the planets' periods uniformly in log-space  ($P_j$ from 1 to 980~d, the time coverage of the NIRPS data) and masses ($M_{p,\,j}$ from 0.3\,M$_\oplus$ to 1.9\,M$_\mathrm{J}$). The stellar mass is drawn from a normal distribution $\mathcal{N}(M_\star,\,\sigma_\star)$ based on the mass $M_\star$ and uncertainty $\sigma_\star$ of \hbox{TRAPPIST-1} \citepalias[$0.0898\pm0.0023$\,M$_\odot$;][]{agolRefiningTransittimingPhotometric2021}. Based on the low dispersion of mutual inclinations among \textit{Kepler} multiplanet systems \citep{ballardKEPLERDICHOTOMYDWARFS2016, daiLargerMutualInclinations2018}, we assumed nearly coplanar orbits with the other \hbox{TRAPPIST-1} planets. The inclination is sampled from a normal distribution $\mathcal{N}(i_p,\,\sigma_i)$ with $i_p=89.778\degree$, the median of the planets' inclinations \citepalias{agolRefiningTransittimingPhotometric2021}, and $\sigma_i=2\degree$ \citep{ballardKEPLERDICHOTOMYDWARFS2016}. The orbital phase is sampled uniformly from 0 to $2\pi$\,rad and the orbit is assumed to be circular ($e=0$). Once the signal is injected, we recomputed the activity model for each dataset, using the same GP hyperparameters as the best-fit model, and subtracted it from the time series \citep{gonzalezhernandezSubEarthmassPlanetOrbiting2024}.

\begin{figure}[t]
\centering
\includegraphics[width=\linewidth]{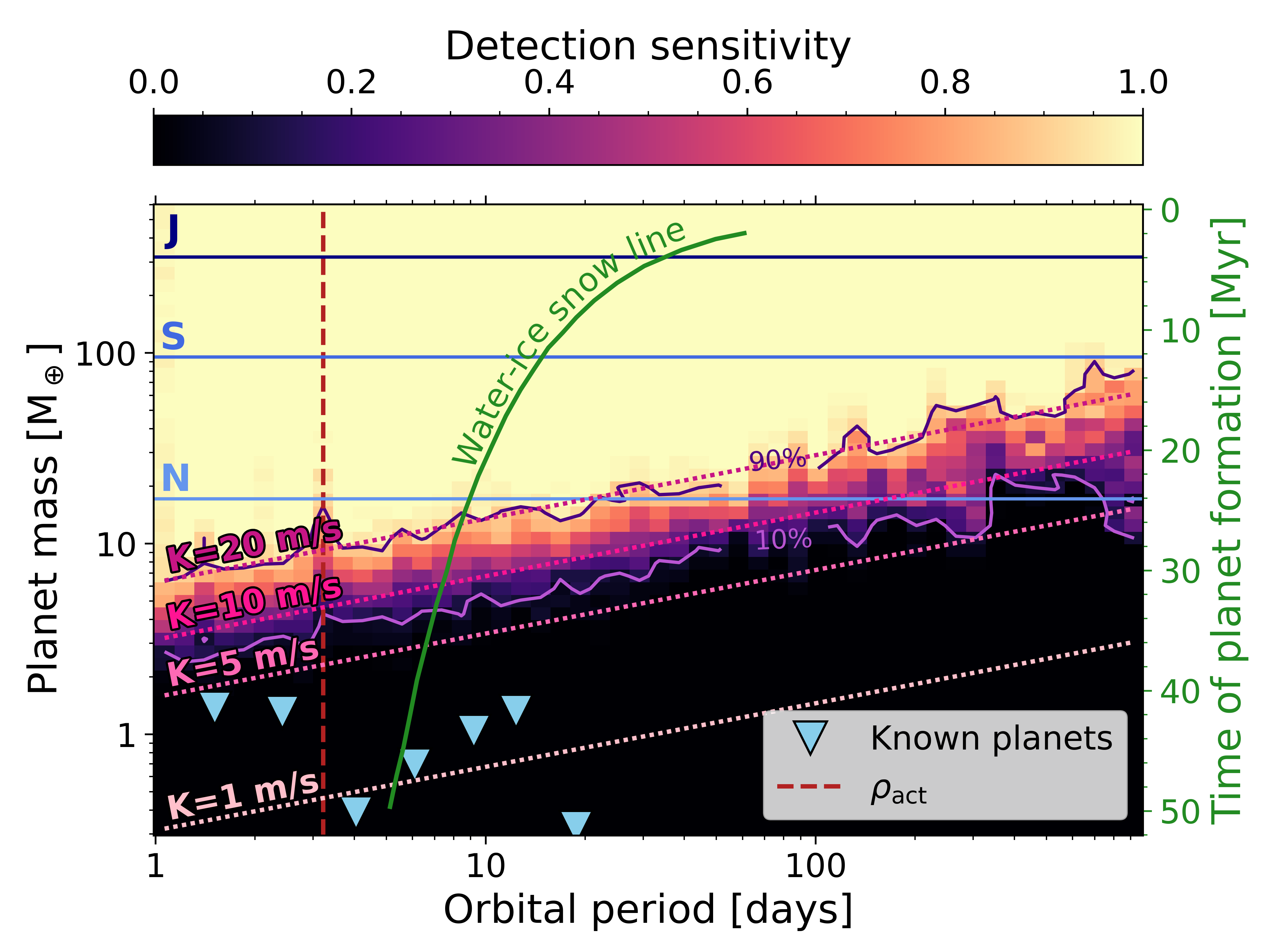}
\caption{Sensitivity of the NIRPS RV measurements to the presence of additional planets in the \hbox{TRAPPIST-1} system as a function of planet mass and orbital period. The detection sensitivity, indicated by the colorbar, is computed for each $10\times10$ bins in the parameter space, ranging from 0 (black) to 1 (pale yellow). Medium and dark purple contours mark the 10\% and 90\% sensitivity limits, respectively. The known \hbox{TRAPPIST-1} planets are shown with the blue triangles while the vertical dashed red line highlights the stellar activity period found in the RV data, $\rho_\mathrm{act}$. The masses of Jupiter (J), Saturn (S) and Neptune (N) are indicated with the horizontal solid blue lines. RV semi-amplitudes of 1, 5, 10 and 20\,m\,s$^{-1}$ are marked by the pink dotted lines. Following \citet{unterbornInwardMigrationTRAPPIST12018}, we show the position of the water-ice snow line evolving through time in the protoplanetary disk, assuming a condensation temperature of 212\,K. Depending on the time of planet formation with regards to the star (rightmost $y$ axis), the water-ice snow line will correspond to shorter orbital periods (green line) as the stellar luminosity decreases with time. Any planet more massive than Saturn would have been detected by NIRPS out to 2.7~yr in orbital period.}
\label{fig:detectionmapNIRPS}
\end{figure}

To determine if a signal is recovered, we used criteria from \citet{cloutierMorePreciseMass2021} and \citet{cherubimTOI1695WaterWorld2023}. Firstly, there must be a peak in the normalized Generalized Lomb-Scargle periodogram (GLS; \citealt{zechmeisterGeneralisedLombScarglePeriodogram2009}) with FAP$\leq1$\% within 10\% of the injected period $P_j$. Then, we computed the Bayesian Information Criterion (BIC) to verify that the Keplerian model is strongly favored over a flat line, checking that $\Delta\mathrm{BIC}=\mathrm{BIC}_\mathrm{flat}-\mathrm{BIC}_\mathrm{kep}\geq10$. The $\mathrm{BIC}=-2\ln{\mathcal{L}}+\nu\ln{N}$ is computed from $\mathcal{L}$, the likelihood of the RV data given the model, $\nu$, the number of parameters in the model ($\nu=6$ for the Keplerian and $\nu=1$ for the flat line), and $N$, the number of RV measurements. The planet recovery is considered successful if the two criteria are met. 

The RV sensitivity is computed as the ratio of recovered planets to the number of injected planets, represented in Fig.~\ref{fig:detectionmapNIRPS} as a function of planet mass and orbital period for bins of $10\times10$ in the $(\log M_{p,\,j},\,\log P_j)$ parameter space. While the NIRPS RV measurements are not precise enough to detect lower-mass objects such as the seven known \hbox{TRAPPIST-1} planets or even sub-Neptune candidates, they place strong constraints on the presence of giant planets: with $>90\%$ recovery, Jupiter- and Saturn-mass planets on $<2.7$-yr orbits would have been detected by NIRPS. Since we do not see these signals in our time series, we can exclude the presence of such planets from the TRAPPIST-1 system. The stellar activity period ($\rho_\mathrm{act}$, modeled by the GP) only slightly limits the detection sensitivity around 3.22~d for Neptune-mass planets. However, we do not expect the presence of a massive planet in the middle of the \hbox{TRAPPIST-1} planetary system to have gone unnoticed by transits and TTV analyses. Therefore, we can exclude the presence of planets more massive than Saturn out to 2.7-yr orbits and Neptune-mass planets out to 20~d. These new limits significantly add to previous astrometric constraints which excluded $\gtrsim4.6$\,M$_\mathrm{Jup}$ planets on one-year orbits and $\gtrsim1.6$\,M$_\mathrm{Jup}$ planets on five-year orbits \citep{bossAstrometricConstraintsMasses2017}.

Following \citet{unterbornInwardMigrationTRAPPIST12018}, we considered the position of the water-ice snow line as it evolves through the life of the protoplanetary disk, assuming a condensation temperature of 212\,K owing to the greater surface density of M dwarf disks compared to the solar nebula (170\,K). In Fig.~\ref{fig:detectionmapNIRPS}, we show the position of the snow line depending on when the planet forms inside the disk, with a maximal orbital period of 60~d which decreases with time as the star becomes less luminous. \citet{raymondUpperLimitLate2021} demonstrate that the \hbox{TRAPPIST-1} system must have formed within a few million years to preserve the planetary resonant chain. This timescale points to the water-ice snow line being positioned between 17 and 60~d during planet formation. The NIRPS RVs find no evidence for the presence of gas giant planets inside or beyond the snow line ($M_p>M_\mathrm{Saturn}$ out to 2.7-yr orbits), in agreement with predictions from \citet{ormelFormationTRAPPIST1Other2017}, who hypothesized that close-in compact systems such as \hbox{TRAPPIST-1} would rarely harbor distant giant planets. Such massive planets would form rapidly at the snow line, stopping the flux of pebbles to the inner disk and quenching it from planet-building material. 

Still, with the available data, we cannot exclude the possibility of a Neptune or sub-Neptune-mass planet beyond the snow line. Placing definitive constraints on the presence of such planet would require reaching sufficient precision to detect $\sim$5\,m\,s$^{-1}$ RV semi-amplitudes ($K$, see pink dotted lines in Fig.~\ref{fig:detectionmapNIRPS}) on long timescales. With the data currently available, we expect to retrieve approximately 90\% of signals with $K=20$\,m\,s$^{-1}$. For NIRPS to constrain Neptune-mass planets beyond the snow line in the \hbox{TRAPPIST-1} system, we would require a two-fold improvement in the sensitivity of the RV data for orbital periods under 100~d and a four-fold improvement for periods out to 2.7~yr. As the amplitude of the signal would be comparable to the level of stellar activity, any survey dedicated to this search should employ a careful observing strategy to simultaneously characterize the stellar activity on short timescale and the presumptive long-period planet.

\section{Conclusion}\label{sec:conclusion}

This work presents a comprehensive RV analysis of the SPIRou and NIRPS near-infrared RV measurements obtained for the \hbox{TRAPPIST-1} system between 2019 and 2025. We first characterized the stellar activity by deriving the rotation period from the \textit{K2} ($P_\mathrm{rot}=3.302\pm0.005$~d) and \textit{TESS} ($P_\mathrm{rot}=3.6\pm1.0$~d) photometric light curves. From the RV measurements, we find a period of $3.22^{+0.22}_{-0.20}$~d. The consistency of the recovered value across multiple time series spanning $>6.5$~yr confirms the rotation of the star as the source of the periodicity.

We also used polarimetric sequences from SPIRou to place an upper limit on the large-scale magnetic field. The non-detection of the longitudinal field ($|B_l|<40$\,G; 3$\sigma$) excludes an inclined dipolar field of polar strength $>0.1$\,kG. The remaining possible geometries, either a strong axisymmetric dipole or a weak multipolar field, are both consistent with previous reports on magnetic topologies for rapidly rotating late-type M dwarfs. As even a slight misalignment of the dipole and the stellar rotation axis would lead to a detectable $B_l$ and considering the short activity timescale seen with RV, the multipolar field configuration is favored.

From the SPIRou and NIRPS RV measurements, we detected the planetary signal of the \hbox{TRAPPIST-1} system. While the current RVs are not precise enough to allow the individual detection of the seven planets, we used the relative masses from the TTV analysis to inform an RV model that combines the signal of all the planets, detecting them together by adjusting the total amplitude of the planetary RV signal through TRAPPIST-1\,b. We find $K_{\mathrm{b}}=3.65^{+0.78}_{-0.83}$\,m\,s$^{-1}$, corresponding to $M_{p,\,\mathrm{b}}=1.31\pm0.29$\,M$_\oplus$, in agreement with the mass obtained from TTVs \citepalias[$1.374\pm0.069$\,M$_\oplus$;][]{agolRefiningTransittimingPhotometric2021}. Thus, we demonstrate for the first time that that RV observations of \hbox{TRAPPIST-1} are consistent with the TTV model of the system.

From the systematics-free NIRPS RV measurements, we constrained the presence of giant planets in the \hbox{TRAPPIST-1} system. We exclude planets more massive than Saturn out to 2.7-yr orbits and Neptune-mass planets out to 20~d, tightening previous constraints from astrometric measurements. Constraining a Neptune-mass planet beyond the water-ice snow line would require a factor two to four improvement in the precision of the NIRPS RVs, which could be achieved through longer exposure times in a dedicated survey while monitoring the stellar activity on short timescale.

\section*{Data availability}

The TRAPPIST-1 photometric time series used in this work are available through GitHub (\url{https://github.com/rodluger/trappist1}) and MAST (\url{http://archive.stsci.edu/tess/}) for \textit{K2} and \textit{TESS}, respectively. The raw and reduced SPIRou data are publicly available through the CADC (\url{https://www.cadc-ccda.hia-iha.nrc-cnrc.gc.ca}). We provide the per-night SPIRou and NIRPS heliocentric RV measurements of TRAPPIST-1 analyzed in this work in Table~\ref{tab:RVmeasurements}, with the full content being available in electronic form at the Centre de Données astronomiques de Strasbourg (CDS) via anonymous ftp to cdsarc.u-strasbg.fr (130.79.128.5) or via \url{http://cdsweb.u-strasbg.fr/cgi-bin/qcat?J/A+A/}. The full posterior distributions of the of the different models (summarized in Fig.~\ref{fig:KbEvidenceAllModels}) are available in a Zenodo repository (\url{https://doi.org/10.5281/zenodo.21844190}).

\begin{acknowledgements}
      We thank the anonymous referee for valuable comments that improved the quality of this paper.
      
      This work is based on observations obtained at the Canada-France-Hawai'i Telescope (CFHT) which is operated by the National Research Council of Canada, the Institut National des Sciences de l'Univers of the Centre National de la Recherche Scientifique of France, and the University of Hawai'i. CFHT is located on Maunakea on Hawai'i Island, a mountain of considerable cultural, natural, and ecological significance. Maunakea is a sacred site to Native Hawaiians, also known as Kānaka 'Ōiwi. Quality observations are made possible by relentless effort of the entire staff at Canada-France-Hawai'i Telescope. Based on observations obtained with SPIRou, an international project led by Institut de Recherche en Astrophysique et Planetologie, Toulouse, France.
      
      This paper includes data collected by the \textit{Kepler} mission and obtained from the MAST data archive at the Space Telescope Science Institute (STScI). Funding for the \textit{Kepler} mission is provided by the NASA Science Mission Directorate. STScI is operated by the Association of Universities for Research in Astronomy, Inc., under NASA contract NAS 5–26555.

      This paper includes data collected with the \textit{TESS} mission, obtained from the MAST data archive at the Space Telescope Science Institute (STScI). Funding for the \textit{TESS} mission is provided by the NASA Explorer Program. We acknowledge the use of public \textit{TESS} Alert data from pipelines at the \textit{TESS} Science Office and at the \textit{TESS} Science Processing Operations Center. Resources supporting this work were provided by the NASA High-End Computing (HEC) Program through the NASA Advanced Supercomputing (NAS) Division at Ames Research Center for the production of the SPOC data products. This paper includes data collected by the \textit{TESS} mission that are publicly available from the Mikulski Archive for Space Telescopes (MAST). 
      
      AL  acknowledges support from the Fonds de recherche du Qu\'ebec (FRQ) - Secteur Nature et technologies under file \#317615 and \#349961. This work was supported by grant \url{https://doi.org/10.69777/377645} from the FRQ and by the Center for research in astrophysics of Québec (AstroQuébec). AL, RD, CC, OL, \'EA, NJC, LMo, RA, FBa, BB, PL, LMa \& JPW  acknowledge support from the Trottier Family Foundation and the Trottier Institute for Research on Exoplanets. RD, \'EA, FBa \& LMa  acknowledges support from Canada Foundation for Innovation (CFI) program, the Universit\'e de Montr\'eal and Universit\'e Laval, the Canada Economic Development (CED) program and the Ministere of Economy, Innovation and Energy (MEIE). E.A. acknowledges NASA XRP 80NSSC21K1111 and ICAR 80NSSC23K1398. This work has been carried out within the framework of the NCCR PlanetS supported by the Swiss National Science Foundation under grants 51NF40\_182901 and 51NF40\_205606. XB, XDe, AC \& TF  acknowledge funding from the French ANR under contract number ANR\-24\-CE49\-3397 (ORVET), and the French National Research Agency in the framework of the Investissements d'Avenir program (ANR-15-IDEX-02), through the funding of the ``Origin of Life" project of the Grenoble-Alpes University. CM \& J-FD  acknowledge funding from the European Research Council (ERC) under the H2020 research \& innovation programme (grant agreement \#740651 New-Worlds). LMo  acknowledges the support of the Natural Sciences and Engineering Research Council of Canada (NSERC), [funding reference number 589653]. RA  acknowledges the Swiss National Science Foundation (SNSF) support under the Post-Doc Mobility grant P500PT\_222212. Research activities of the Board of Observational and Instrumental Astronomy at the Federal University of Rio Grande do Norte (NAOS) are supported by continuous grants from the Brazilian funding agency CNPq. This study was financed in part by the Coordena\c{c}\~ao de Aperfei\c{c}oamento de Pessoal de N\'ivel Superior -- Brasil (CAPES) -- Finance Code 001, and by the program CAPES/Print. ICL  acknowledges CNPq research fellowships (Grant No. 313103/2022-4). KAM  acknowledges support from the SNSF under the Postdoc Mobility grant P500PT\_230225. BLCM  acknowledge CAPES postdoctoral fellowships and CNPq research fellowships (Grant No. 305804/2022-7). RC  acknowledges support from the Canada Research Chairs Program and the NSERC. NBC  acknowledges support from an NSERC Discovery Grant, a Canada Research Chair, and an Arthur B. McDonald Fellowship, and thanks the Trottier Space Institute for its financial support and dynamic intellectual environment. EC \& NCS  acknowledge the support from FCT - Funda\c{c}\~ao para a Ci\^encia e a Tecnologia through national funds by these grants: UIDB/04434/2020, UIDP/04434/2020, UID/04434/2025. JRM  acknowledges CNPq research fellowships (Grant No. 308928/2019-9). XDu  acknowledges the support from the ERC under the European Union’s Horizon 2020 research and innovation programme (grant agreement SCORE No 851555) and from the SNSF under the grant SPECTRE (No 200021\_215200). Y.G.C.F. acknowledges funding from the SNSF through project P500PN\_217951. JIGH, AKS \& ASM  acknowledge financial support from the Spanish Ministry of Science, Innovation and Universities (MICIU) projects PID2020-117493GB-I00 and PID2023-149982NB-I00. PL  acknowledges financial support from the Severo Ochoa grant CEX2021-001131-S funded by MCIN/AEI/10.13039/501100011033. PL  is funded by the European Union (ERC, THIRSTEE, 101164189). Co-funded by the European Union (ERC, FIERCE, 101052347). Views and opinions expressed are however those of the author(s) only and do not necessarily reflect those of the European Union or the European Research Council. Neither the European Union nor the granting authority can be held responsible for them. E.M. acknowledges funding from FAPEMIG under project number APQ-02493-22 and a research productivity grant number 309829/2022-4 awarded by the CNPq, Brazil. CP  acknowledges support from the NSERC Vanier scholarship, and the Trottier Family Foundation. CP also acknowledges support from the E. Margaret Burbidge Prize Postdoctoral Fellowship from the Brinson Foundation. AKS  acknowledges financial support from La Caixa Foundation (ID 100010434) under the grant LCF/BQ/DI23/11990071. GAW is supported by a Discovery Grant from the NSERC.
\end{acknowledgements}

\bibliographystyle{aabib}
\bibliography{zotero_updated}

\begin{appendix}
\nolinenumbers

\section{The impact of image persistence on SPIRou and NIRPS spectra}\label{sec:persistence}

Like all instruments with IR detectors, SPIRou and NIRPS face the non-trivial problem of image persistence, or simply persistence. It is thought that persistence arises from defects in the detector diodes: charges are trapped and then slowly released over time, creating an afterimage that can be detected in subsequent exposures \citep{smithCalibrationImagePersistence2008, smithTheoryImagePersistence2008, longPersistenceWFC3IR2015}. The persistence image consists of a decaying remnant of all the previous observations. Its amplitude is related to the accumulated flux at the time of detector reset and decreases as the inverse of time \citep{artigauH4RGCharacterizationHighresolution2018, cookAPEROPipelinEReduce2022}. Persistence has been known to limit data quality of instruments with mercury, cadmium, tellurium (HgCdTe) IR detectors (e.g., HST/WFC3/IR; \citealt{Long2010WFC3IRPA}), with significant impact on faint targets for which the relative amplitude of the image persistence is more important. For SPIRou, $H >$ 10\,mag is cited as the regime where persistence can have a significant impact on the RV estimate \citep{donatiSPIRouNIRVelocimetry2020}, including notably \hbox{TRAPPIST-1} ($H=10.7$\,mag).

\begin{figure}[h]
    \centering
    \includegraphics[width=\linewidth]{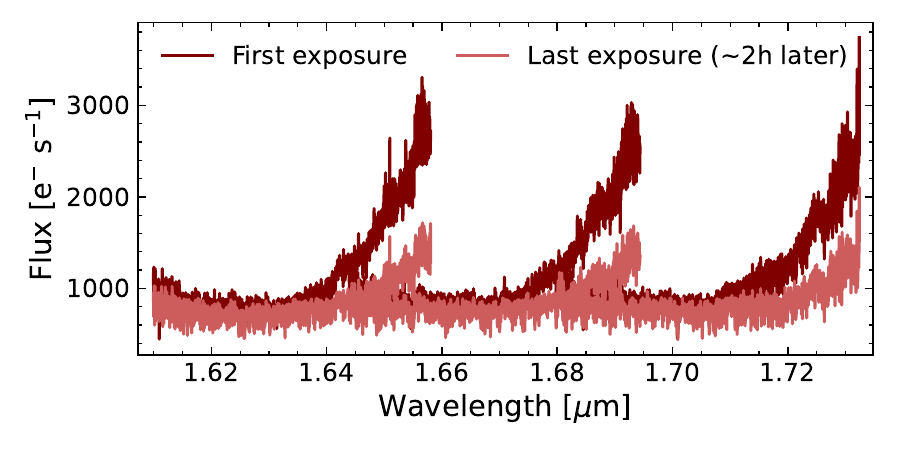}
    \caption{Flux (in electrons/s) of three SPIRou orders for the first (dark red) and the last (light red) exposures of the \hbox{TRAPPIST-1} observation of September 25, 2019. The two exposures (512~s each) are separated by $\sim2$h, being part of the baseline of a transit observation. Right before TRAPPIST-1, an observation at S/N$\sim$150 of a bright object (GJ~752~A, $H=4.9$\,mag) was performed. Persistence has a strong asymmetrical shape on the SPIRou detector, as shown by the large flux excess for the part of the orders covering longer wavelengths. With time, the trace spectrum of the previously observed star gradually fades away, but persists for many hours.}
    \label{fig:persistenceOrders}
\end{figure}

The persistence response differs from pixel to pixel and is unique to each detector. Even with SPIRou and NIRPS both having a $4096\times4096$ pixel HAWAII-4RG (H4RG) IR array, built in the same batch by Teledyne Scientific \& Imaging, their observed persistence images vary drastically. In particular, the persistence image on the SPIRou detector displays strong spatial variations. As shown in Fig.~\ref{fig:persistenceOrders}, we see a clear excess of flux on the longer-wavelength side of each of the \'{e}chelle orders. This excess is due to the persistence image of preceding observations, including a relatively bright high-S/N target (GJ~752~A, $H=4.9$\,mag, S/N$\sim$150) right before the \hbox{TRAPPIST-1} observation. The excess flux decreases with time, but is still present after 2h. For a faint star like TRAPPIST-1, the image persistence of a high S/N signal, likely from a bright star, can contribute more than 50\% of the total flux in the worst regions of the SPIRou detector (Fig.~\ref{fig:persistenceOrders}).

\begin{figure*}[ht]
    \centering
    \sidecaption
    \includegraphics[width=12cm]{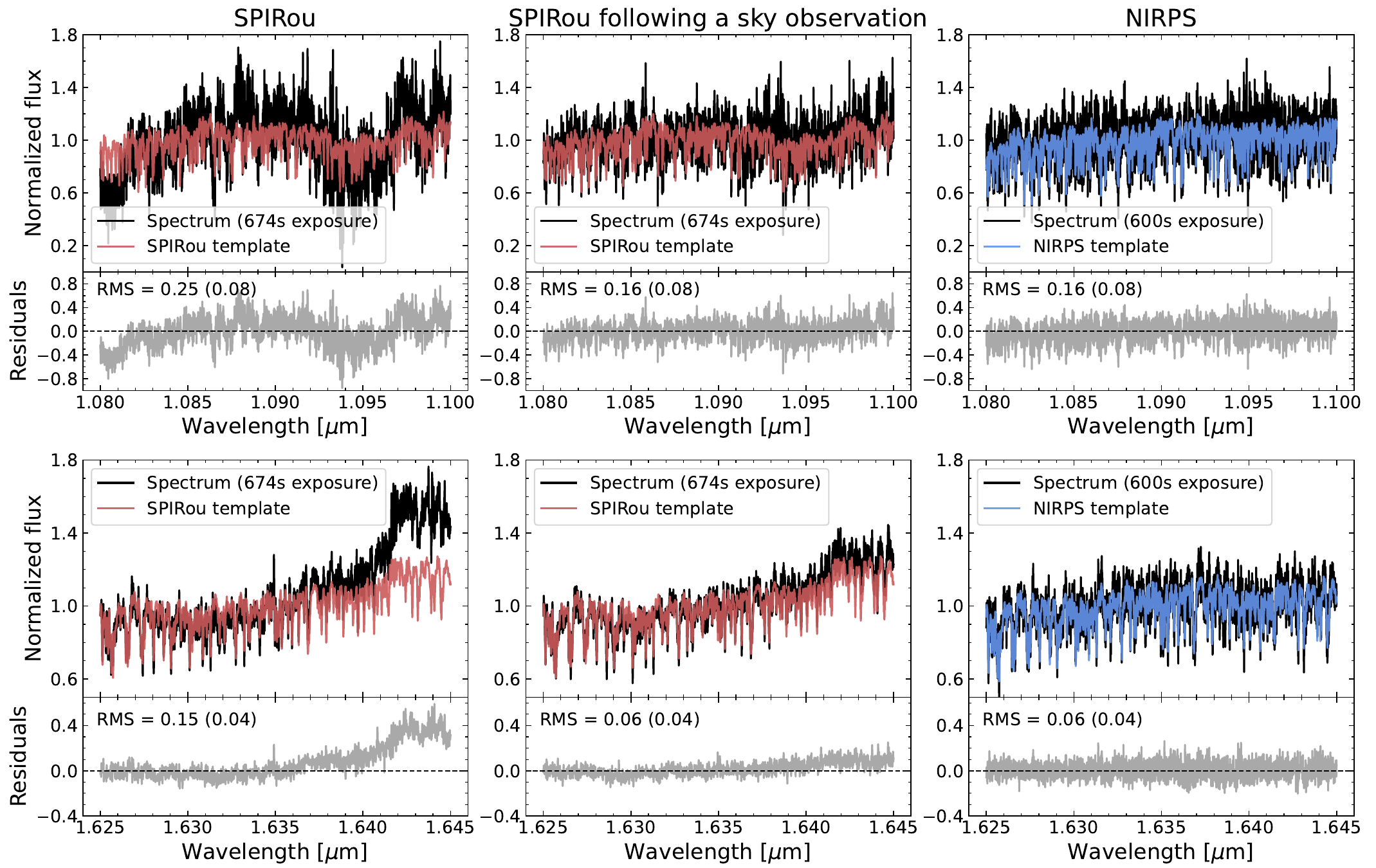}
    \caption{Individual telluric-corrected \hbox{TRAPPIST-1} spectra (black line) compared to the stellar template constructed from all the observations obtained with SPIRou (red line) and NIRPS (blue line). The residuals are shown in gray in the lower panel of each subplot, with their calculated root mean square (RMS). The expected RMS from the S/N of the spectra is written in parentheses. Three cases are highlighted: \textit{(left column)} a SPIRou observation made right after a high-S/N bright target (GJ~849, S/N$\sim$100, $H=5.9$\,mag), \textit{(middle column)} a SPIRou observation made following a 250\,s sky observation after a high-S/N bright target (GJ~846, S/N$\sim$100, $H=5.6$\,mag) and \textit{(right column)} a NIRPS observation of similar exposure time following a 300\,s sky after a similarly high-S/N and bright target (HR~7830, S/N$\sim$160, $H=5.8$\,mag). Each row presents a different spectral region. The persistence effect is strongest in the first case, distorting the spectrum and leading to sub-optimal telluric correction.}
    \label{fig:templateVSspectra}
\end{figure*}

When stitching the SPIRou orders together, the strong flux asymmetry resulting from the image persistence distorts the shape of the stellar spectrum. In Fig.~\ref{fig:templateVSspectra}, we show this distortion for two spectral regions, comparing individual telluric-corrected \hbox{TRAPPIST-1} spectra to the stellar template constructed from all the telluric-corrected SPIRou or NIRPS observations. In the first column, we show a case similar to that presented in Fig.~\ref{fig:persistenceOrders}, with an exposure of \hbox{TRAPPIST-1} following a relatively high-S/N observation (S/N$\sim$100) of a bright target (GJ~849, $H=5.9$\,mag). The spectrum is heavily distorted compared to the template, especially at the junction of spectral orders ($\sim1.080,\,1.095$ and $1.645\,\mu$m). 

In practice, even the SPIRou template used as comparison contains persistence. Because of its structured spatial distribution, the persistence is not averaged out in the template, even if its amplitude is varying from one observation to the next. For example, the SPIRou template exhibits a small kink near $1.095\,\mu$m and a slope longward of $1.64\,\mu$m, at the junctions of orders. These shapes are not present in the NIRPS template, highlighting their origin as persistence-driven rather that stellar. Indeed, NIRPS does not have the same order junctions as SPIRou and the persistence in its array is both of lower amplitude and less spatially variable. We can thus interpret the SPIRou template as the average persistence level across all \hbox{TRAPPIST-1} observations.

The middle column of Fig.~\ref{fig:templateVSspectra} shows a similar case as the left column: the first exposure of \hbox{TRAPPIST-1} following a bright target (GJ~846, $H=5.6$\,mag) with a S/N$\sim$100. However, this time, a 250\,s sky was observed in between, allowing for an additional detector reset and for the amplitude of the persistence to decrease. The spectrum is therefore much less distorted, with persistence near the average level of the template. The residuals are still structured, but their root mean square (RMS) is closer to what is expected from the S/N of the spectrum (indicated in parentheses in the lower panels of Fig.~\ref{fig:templateVSspectra}). Starting in 2020 ($\sim2\,458\,980$\,BJD), we adopted this strategy of observing a 250\,s sky before every \hbox{TRAPPIST-1} observations with SPIRou, limiting but not removing the persistence seen in the spectra (see the first two columns of Fig.~\ref{fig:templateVSspectra}).

In the right column of Fig.~\ref{fig:templateVSspectra}, we show a NIRPS spectrum obtained for a similar exposure time as the SPIRou observations (600\,s instead of 674\,s). It consists of the first exposure of a transit observation, which are always preceded by a 300\,s sky in NIRPS. The structure of the image persistence on the NIRPS detector is much more uniform than for SPIRou, leading to less spectral distortion. The amplitude of the persistence is also reduced, with a decrease of the fractional persistence flux down to $5\times10^{-5}$ of the previous target's brightness in 2--3~min \citep{bouchyNIRPSJoiningHARPS2025}. In comparison, for SPIRou, the fractional flux decreases to $\sim10^{-4}$ in 2~min, averaged over the whole detector \citep{artigauH4RGCharacterizationHighresolution2018}. Still, the residuals between the spectrum and the NIRPS template display a higher RMS than what is expected for the S/N of the spectrum. This level of disagreement, similar to that of the SPIRou observation following a sky, indicates that additional noises are still present in the spectra.

For a faint target like TRAPPIST-1, the persistence effect causes multiple issues. We have, superimposed on our science spectra, a decaying afterimage of all the stars previously observed during the night. These time-variable ``ghost spectra'' dilute our spectral lines in unpredictable ways from one observation to the next. All spectroscopic indicators that rely on a measurement of the line depth or shape will thus be affected by persistence (dLW \citep{zechmeisterSpectrumRadialVelocity2018}, \textit{d}Temp \citep{artigauMeasuringSubKelvinVariations2024}, $\log{R'_\mathrm{HK}}$ \citep{hartmannAnalysisVaughanPrestonSurvey1984, noyesRotationConvectionMagnetic1984, suarezmascarenoRotationPeriodsLatetype2015}, etc.). Additionally, persistence-driven line dilution severely impacts the fit and removal of telluric features. A telluric correction performed on a spectrum affected by persistence will leave  telluric residuals. For SPIRou in particular, where the spectra are also distorted due to the asymmetric nature of the image persistence, the telluric correction is even more challenging, suggesting even larger telluric residuals. Therefore, even if persistence does not affect the position of the stellar spectral lines, allowing for RV measurements of the system, it leads to telluric residuals which might inject systematic signal in the RVs. The persistence level differs  stochastically from one observation to another, but there are telluric residuals consistently present in the spectra. For SPIRou, the residuals are major enough to cause the quasi-periodic systematic signal seen in the \hbox{TRAPPIST-1} RVs (Sect.~\ref{sec:systematics}). For NIRPS, the lower persistence level leads to a more robust telluric correction and, within the uncertainties, to systematics-free RVs.

\section{Robustness of the planetary detection to different systematics and stellar activity treatments}\label{sec:AppendixRobustness}

To assess the robustness of the RV planetary detection, we explored alternative parameterizations for the systematics and stellar activity components of the full model described in Sect.~\ref{sec:modelRV}. For the SPIRou systematics, we considered the sinusoidal model introduced in Sect.~\ref{sec:systematics} in addition to the baseline model, which uses a GP with SHO kernel. The sinusoidal model is constructed from the sum of two sinusoids with their own amplitudes ($A$) and phases ($\Phi$), but related periods ($P_\mathrm{syst}$ and $P_\mathrm{syst}/2$), for a total of five parameters. In comparison, the GP SHO model has three hyperparameters (see Table~\ref{table:parameters}).

For the stellar activity, we explored an alternative GP model, using a quasi-periodic (QP) kernel instead of the SHO of the baseline model. We adapted the QP kernel of \citet{foreman-mackeyFastScalableGaussian2017}, frequently used to model stellar activity with modeling tools like \texttt{juliet} \citep{espinozaJulietVersatileModelling2019}, to the framework of \texttt{celerite2} \citep{foreman-mackeyScalableBackpropagationGaussian2018}. The QP kernel is defined through four hyperparameters (see equation~56 of \citealt{foreman-mackeyFastScalableGaussian2017}): $B$, the amplitude of the GP, $C$, an additive factor impacting the amplitude, $L$, the timescale and $P$, the rotation period. Compared to the SHO kernel used in the baseline model, the QP kernel has two more hyperparameters.

\begin{figure*}[ht]
    \centering
    \sidecaption
  \includegraphics[width=12cm]{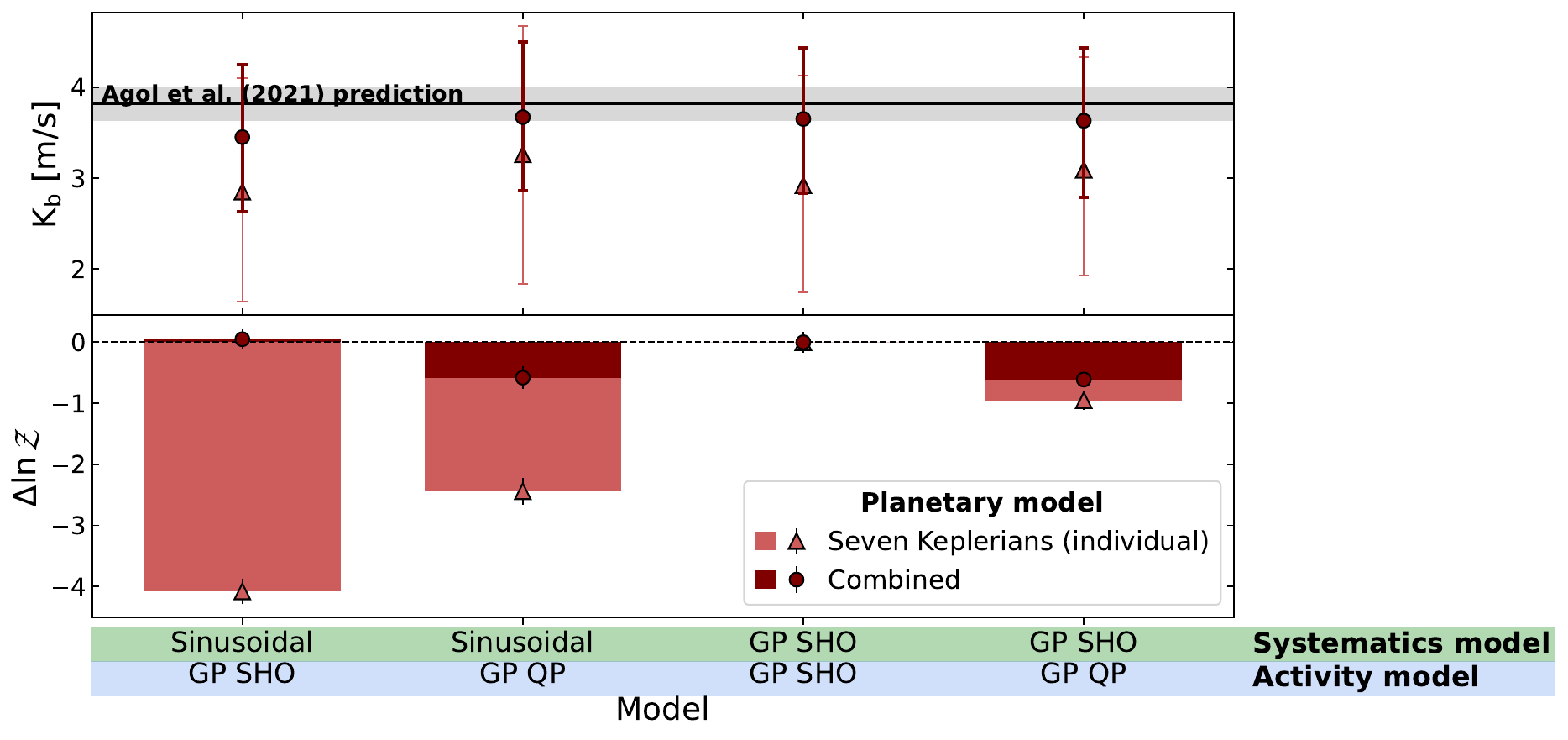}
    \caption{Comparison in the recovered RV semi-amplitude of TRAPPIST-1\,b ($K_\mathrm{b}$, top panel) and the difference in Bayesian log-evidence compared to the baseline systematics and stellar activity model ($\Delta\ln\mathcal{Z}$, bottom panel). The baseline model, which uses a GP with stochastically-driven harmonic oscillator (SHO) kernel to model both systematics and stellar activity, is the most favored set up. All the other configurations, using a sinusoidal model for the systematics and a GP with quasi-periodic (QP) kernel for the stellar activity, have lower or equivalent log-evidence to the baseline model. For each model, we show the $K_\mathrm{b}$ and $\Delta\ln\mathcal{Z}$ for a model fitting for the individual (light red, triangle markers, based on Sect.~\ref{sec:SevenKepPlanetModel}) and combined (dark red, circle markers, based on Sect.~\ref{sec:CombinedPlanetModel}) planetary signals. In all cases, we recover perfectly consistent $K_\mathrm{b}$, in agreement with the prediction of \citetalias{agolRefiningTransittimingPhotometric2021} (black line and gray shaded region, top panel), demonstrating that the RV planet detection is not a result of the chosen systematics and stellar activity model.}
    \label{fig:KbEvidenceAllModels}
\end{figure*}

We fitted all combinations of the two systematics (sinusoidal and GP SHO) and stellar activity (GP QP and GP SHO) models to the SPIRou+NIRPS datasets, using both the seven Keplerian (Sect.~\ref{sec:SevenKepPlanetModel}) and combined (Sect.~\ref{sec:CombinedPlanetModel}) planetary parameterizations. The top panel of Fig.~\ref{fig:KbEvidenceAllModels} shows the measured value for the RV semi-amplitude of TRAPPIST-1\,b ($K_\mathrm{b}$), indicating that all models recover equivalent results which are consistent with the \citetalias{agolRefiningTransittimingPhotometric2021} prediction. In the bottom panel of Fig.~\ref{fig:KbEvidenceAllModels}, we show the difference in Bayesian log-evidence of the models compared to the baseline set-up of using GPs with SHO kernels to model both systematics and stellar activity (described in Sect.~\ref{sec:modelRV}). We see that the baseline model (GP SHO + GP SHO) is the most favored set-up for both an individual and combined planetary parametrization. For the combined planetary model specifically, we find that modeling the systematics with a sinusoidal model yields a equivalent Bayesian evidence to the baseline model. Still, for the same $\ln\mathcal{Z}$, we favor the GP SHO model since it also performs well with the seven Keplerian planetary model.

Being favored by Bayesian evidence, we chose to use GP models with SHO kernels to model both the SPIRou systematics and stellar activity as our baseline model (Sect.~\ref{sec:modelRV}). Still, since we recovered perfectly consistent $K_\mathrm{b}$ in all the different tested models, we conclude that the planetary RV detection of the TRAPPIST-1 system is not an artifact of the chosen systematics and stellar activity model.

\section{Results from the RV model}

\begin{figure*}[h!]
\sidecaption
\centering
\includegraphics[width=12cm]{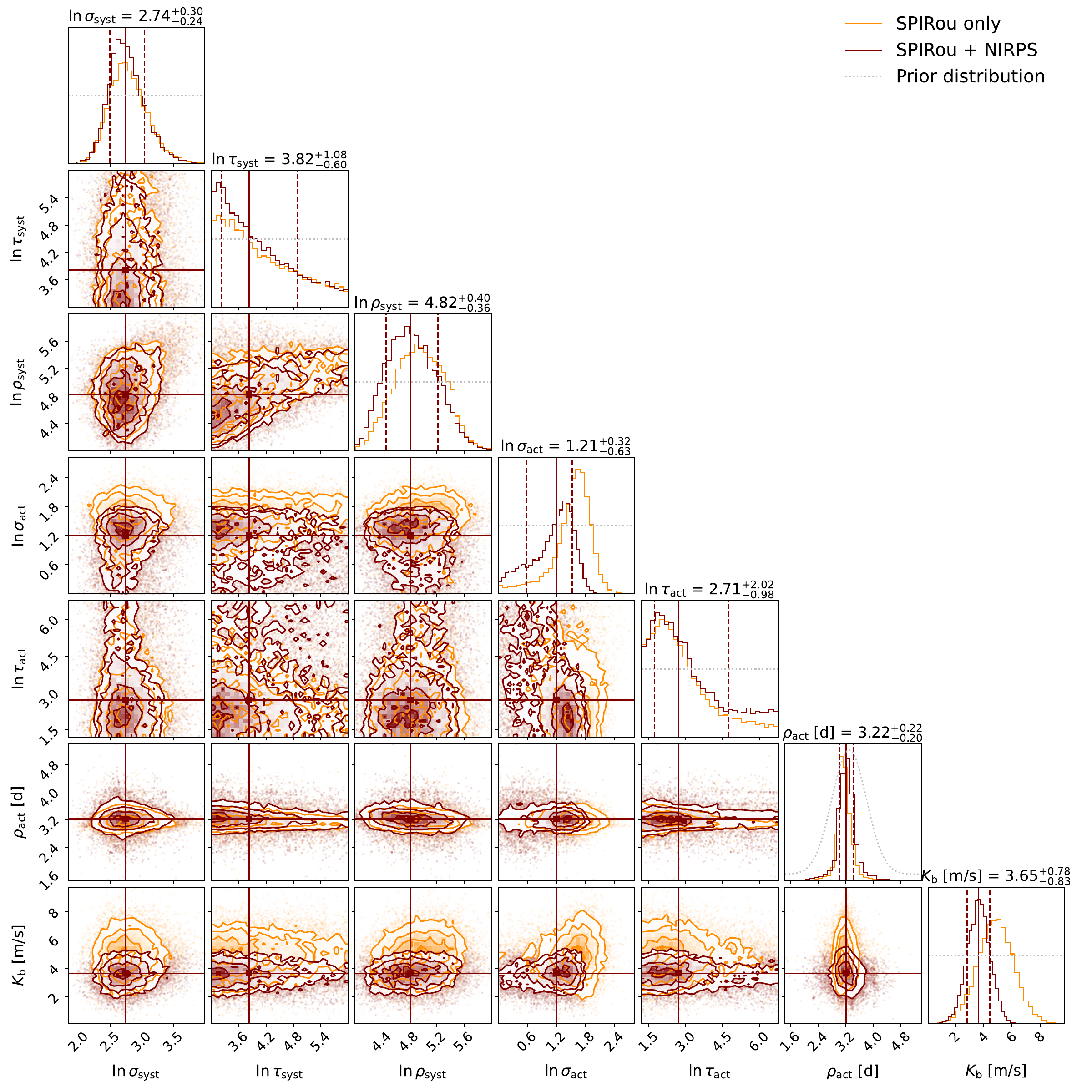}
\caption{Posterior distribution for the best-fit RV model in which we combine the signal of all the \hbox{TRAPPIST-1} planets into $K_{\mathrm{b}}$, the RV semi-amplitude of TRAPPIST-1\,b, through equation~\ref{eq:ratio} (Sect.~\ref{sec:CombinedPlanetModel}). We show the results for fitting the model on only the SPIRou RVs (orange) and on the SPIRou and NIRPS RVs (dark red). The prior distribution used for each parameter is indicated by the gray dotted lines. For the SPIRou+NIRPS model, dashed lines indicate the 16$^{\rm th}$ and 84$^{\rm th}$ percentiles, with a solid line highlighting the median. Note that we only show the parameters of interest, having removed the instrument-specific means and log-jitters ($\phi_\mathrm{SPIRou}$, $\phi_\mathrm{NIRPS}$, $\ln{s_\mathrm{SPIRou}}$ and $\ln{s_\mathrm{NIRPS}}$, see Table~\ref{table:parameters} for their posterior values). The best-fit model is shown in Fig.~\ref{fig:rv}.}
\label{fig:rvcorner}
\end{figure*}

The posterior distributions of the best-fit model on the SPIRou-only and SPIRou+NIRPS RV datasets are presented in Fig.~\ref{fig:rvcorner}. The best-fit model is made up of the baseline systematic and stellar activity model, using a GP with SHO kernel for both, and the planetary model with a parametrization combining the signal of all of the planets (described in Sect.~\ref{sec:CombinedPlanetModel}).

\section{Radial velocity measurements}

The SPIRou and NIRPS per-night radial velocities used in this work are given in Table~\ref{tab:RVmeasurements}

\begin{table*}[h]
\caption{Per-night radial velocity measurements of \hbox{TRAPPIST-1} with SPIRou and NIRPS.}
\label{tab:RVmeasurements}
\centering
\begin{tabular}{cccc}
\hline\hline
BJD - 2\,400\,000 & Heliocentric RV [m\,s$^{-1}$] & $\sigma_{\rm RV}$ [m\,s$^{-1}$] & Instrument\\
\hline
58649.072435556445 & $-53150.75$ & 3.25 & SPIRou\\
58651.10199739924 & $-53150.78$ & 5.65 & SPIRou\\
58654.09849249385 & $-53138.13$ & 4.93 & SPIRou\\
\ldots & \ldots & \ldots & \ldots \\
59912.59189446457 & $-53189.32$ & 6.46 & NIRPS\\
59917.547336518764 & $-53178.66$ & 5.88 & NIRPS\\
59918.5664135986 & $-53194.37$ & 7.05 & NIRPS\\
\ldots & \ldots & \ldots & \ldots \\
\hline
\end{tabular}
\tablefoot{The full content of Table~\ref{tab:RVmeasurements} is available in machine-readable format at the CDS.}
\end{table*}

\end{appendix}

\end{document}